\documentclass{aa}

\usepackage{graphicx}
\usepackage{txfonts}
\usepackage{natbib}
\usepackage{amsfonts}
\usepackage{amssymb}

\usepackage{amstext}

\begin{document}
\title{
Three-dimensional solar active region magnetohydrostatic models \\ and their stability using Euler potentials}

\author{Terradas\inst{1,2}, J., and Neukirch\inst{3}, T.}

\institute{$^1$Departament de F\'\i sica, Universitat de les Illes Balears (UIB),
E-07122, Spain \\ $^2$Institute of Applied Computing \& Community Code (IAC$^3$),
UIB, Spain\\ \email{jaume.terradas@uib.es}\\
$^3$School of Mathematics and Statistics, University of St Andrews, St Andrews, KY16 9SS, UK
}

\abstract{
Active regions (ARs) are typical magnetic structures found in the solar atmosphere. We 
calculate 
several magnetohydrostatic (MHS) equilibrium models that include the effect of a finite plasma-$\beta$ and gravity
and 
that are representative of these  structures in three dimensions. The construction of the models is based on the use of two Euler potentials, $\alpha$ and $\beta$, that represent the magnetic field as ${\bf B}=\nabla \alpha \times \nabla \beta$. The ideal MHS nonlinear partial differential equations are 
solved numerically
using finite elements in a fixed 3D rectangular domain. The boundary conditions are initially chosen to correspond to a  potential  magnetic field (current-free) with known analytical expressions for the corresponding Euler potentials. The distinctive feature is that we incorporate the effect of shear by progressively deforming the initial potential magnetic field. This procedure is quite generic and allows us to 
generate
a vast variety of MHS models. The thermal structure of the ARs is incorporated through the dependence of gas pressure and temperature 
on 
the Euler potentials. Using this method we achieve the characteristic hot and over-dense plasma 
found in
ARs, but we demonstrate that the method can
also
be applied to study configurations 
with
open magnetic field lines.
Furthermore, we
investigate basic topologies that include neutral lines. Our focus is on the force balance of the structures and we do not consider the energy balance in the constructed models. In addition, we address the 
difficult
question of the stability of the 
calculated 
3D models. We find that if the plasma is 
convectively
stable, then the system is not prone in general to develop magnetic Rayleigh-Taylor instabilities. 
However, when the plasma-$\beta$ is increased or the density at the core of the AR is high then the magnetic configuration becomes unstable due to magnetic buoyancy.
}


\titlerunning{Solar AR MHS models and their stability in 3D}
\authorrunning{Terradas \& Neukirch}

\maketitle

\section{Introduction}\label{sectintro}

It is well established that the structure and dynamics of the solar corona is dominated by the magnetic field. In many structures of the corona, such as active regions (ARs) and coronal holes (CHs), magnetic forces are prevailing and plasma pressure gradients and gravity are often ignored. This is only valid as a first order approximation and leads to the so called force-free field models. Even under this assumption sophisticated numerical computations are required to calculate such force-free fields in three dimensions using as boundary conditions the obtained magnetic field vector measured in the solar photosphere. The reader is referred to \citet{wiegelmannsakurai2021} for a review about this topic. In other regions of the solar atmosphere such as at the interface region between the solar photosphere and corona the relative importance of magnetic and plasma forces change by several orders of magnitude. \citet{zhuwiegelmann2018} have focused on this problem and have solved the magnetohydrostatic (MHS) equations with the help of an optimisation principle. Other approaches have recently applied by \citet{zhuwiegelmann2022}.  A recent review of the use of 3D MHS methods for solar magnetic field extrapolation has been given by
\citet{zhuetal2022}.

Although the assumption of zero plasma beta in the solar corona is commonly applied, it is interesting to assess the possible effects of plasma pressure and gravity on the magnetic field. In particular, and from the practical point of view it is appealing to construct MHS models using methods that deviate from the current trends based on optimisation processes, relaxation techniques or Grad-Rubin methods \citep[see][]{wiegelmannsakurai2021}. The idea of the present paper is to use a method developed and applied in the past but that unfortunately has not been extended to the 3D case at least in the study of coronal structures. As a previous step in two dimensions \citet{terradasetal2022} have recently obtained MHS equilibrium solutions that represent CHs and ARs. Based on the  works of \citet{low1975,low1980} and using the flux function, \citet{terradasetal2022}  have reproduced
the main features of ARs, paying particular attention
to the high pressure and diffuse background of these structures
instead of the single-loop structures. The aim of the present paper is to extend the previous two-dimensional work to the three-dimensional case with the purpose of having a better understanding of the effect of gas pressure and gravity on a more realistic magnetic field geometry. For simplicity reasons the analysis of the energetics of the system due to the presence of conduction, radiation and heating is not considered in the present work. In addition, we mostly focus on closed magnetic states representative of ARs in the solar corona.

In the present work the extension of \citet{terradasetal2022}  to three dimensions is based on the use of Euler potentials (EPs hereafter) instead of the flux function. They were originally devised by Euler to describe incompressible velocity fields. The application of EPs is not new in magnetohydrodynamics (MHD), they are well known mostly in the context of magnetospheric studies \citep[see e.g.][]{cheng1995,zahariaetal2004,zaharia2008}. The reader is referred to the fundamental works on the topic of \citet{stern1967,stern1970} \citep[see also][]{stern1976,stern1994a,stern1994b}. A significant number of examples using EPs can be found in \citet{schindler2006} and also in studies related to magnetic reconnection \citep[][]{hesseschindler1988,hessebirn1993}. The EPs are also referred as Clebsch variables \citep[e.g.][]{roberts1967}.  Due to reasons that are discussed in more detail later, there
is only a limited number of investigations that 
have used EPs to study magnetic structures 
in the solar atmosphere.
\citet{barnessturrock1972} took a model to represent the magnetic-field configuration of a sunspot of one polarity surrounded by a magnetic-field region of opposite polarity and used EPs to study how a force-free field structure can be metastable and converted into
an open field structure by an explosive MHD instability.  \citet{zwingmann1984,zwingmann1987} used EPs to investigate the onset mechanisms of eruptive processes in the solar corona, while \citet{romeouneukirch1999,romeouneukirch2001, romeouneukirch2002}, mainly using 2D or 2.5D structures, have investigated sequences of magnetostatic equilibria that may contain bifurcation points using a similar approach as in \citet{zwingmann1987} and \citet{plattneukirch1994} by employing a numerical continuation method to capture the different branches of the solutions. 
As far as we know, the fully 3D case using EPs has not
been addressed in the analysis of magnetic configurations of the solar
corona and this is one of the main purposes of the present work. 

The general 3D MHS problem is quite intricate and 
analytical solutions are only obtained under very specific conditions \citep[see][]{low1985,low1991,low1992,low1993a,low1993b,neukirch1995,neukirch1997,neukirchrastatter1999, neukirchwiegelmann2019}. The previous works are not based on EPs and essentially assume a very particular form of the current density in order to achieve analytical solutions. EPs allow us a more general treatment of the problem but the drawback is that first, a purely numerical treatment is required in most of the cases, and second the representation of a genuinely 3D magnetic field by two EPs exists as a global representation, valid in the whole domain, only if the magnetic field has a simple topology.  In particular, we can always find Euler potentials $\alpha$ and $\beta$ which represent the magnetic field correctly locally but in 3D
we can only guarantee that the same Euler potentials represent the magnetic field everywhere, if the domain
contains one surface which each field line intersects only once and if the
magnetic field does not have any null points ($\bf B$ = 0) inside the domain, or
if the magnetic field has a vector potential $\bf A$ for which ${\bf A} \cdot {\bf B} = 0$ \citep[][]{rosneretal1989}.

It is known that numerical methods based on finite elements are a powerful route to calculate equilibrium solutions under quite broad conditions. They have successfully been used in the past in 2D by
\citet{zwingmannetal1985,zwingmann1987,plattneukirch1994,romeouneukirch1999,romeouneukirch2001,romeouneukirch2002}. This is the technique chosen in the present work to construct a wide range of MHS solutions in 3D based on EPs. But moving to 3D is challenging for several reasons. First of all the size of the matrices involved in the finite element discretisation increases significantly, slowing down the process of obtaining a solution. Second, an appropriate starting point or seed of the initial distribution of the EPs in 3D is required. For this reason, we still need to use potential magnetic fields with known analytical expressions for the corresponding EPs as a starting point of our finite element calculations. Even if we know the magnetic field components the calculation of the EPs is not straightforward. The initial 
states
are the key ingredient to include more realistic effects, such as magnetic shear, in the configurations. This is achieved by gradually 
changing
the magnetic field (i.e., the EPs) 
on 
the boundaries of the domain. Therefore, a difficulty of the method used here relies on the initial 
configuration
required by the numerical method to converge and achieve a final solution. It is worth mentioning that the goal of this paper is not to use an observed magnetic field to construct a MHS solution. The main objective is to assess the benefits and difficulties of using EPs in 3D 
under 
somewhat
idealised 
conditions.
  
Finally, an important question to be decided is whether the 3D equilibrium configurations that are numerically obtained are in fact stable. Due to the interplay between magnetic and buoyant forces,  
the
magnetic Rayleigh-Taylor
instability
or 
the
Parker’s 
instability
may be present in the system,
which is important information to be able to assess the 
relevance of the model for representing solar magnetic field configurations. 
The presence of electric currents can also affect the stability of the system. In the magnetospheric context MHD eigenmodes and the calculation of field line resonances in 3D has been addressed in the past \citep[see for example][]{cheng1995,cheng2003,rankinetal2006,kabinetal2007}. Here, instead of calculating the eigenmodes of the configuration, we use the result devised by \citet{zwingmann1984,zwingmann1987} in the context of the study of coronal magnetic structures. This author realised that under some conditions the problem of the stability in 3D, although he applied the method to 2.5D, can be reduced to the analysis of the discretised version of an operator that is needed during the calculation of the equilibrium solutions. We apply this procedure to the 3D case in the last part of this work to address the significant issue of the stability of the seized MHS solutions.

\section{Magnetohydrostatic equilibrium in three-dimensions using Euler potentials}\label{sec:mhs}

We look for solutions to the following equation
\begin{align}
    \frac{1}{\mu_0}(\nabla\times {\bf B})\times {\bf B}-\nabla p +\rho {\bf g} =0,\label{eq:mhdstatic}
\end{align}
where $\bf B$ is the magnetic field, $p$ is the gas pressure, $\rho$ the plasma density, $g$ the 
gravitational
acceleration on the solar surface and $\mu_0$ the magnetic permeability of free space. The magnetic field from Maxwell's equations has to satisfy that
\begin{align}
\nabla \cdot {\bf B}=0.\label{eq:divb}
\end{align}
We suppose that the plasma is composed of fully ionised hydrogen that satisfies the ideal gas law
\begin{align}
p=\frac{\mathcal{R}}{\bar{\mu}} \rho  T,\label{eq:ideal}
\end{align}
where $T$ is the temperature, $\mathcal{R}$ the gas constant, and $\bar{\mu}$ the mean atomic weight. The aim is to obtain solutions to the previous equations but we have a system of five equations (Eqs.~(\ref{eq:mhdstatic})-(\ref{eq:ideal})) but six unknowns, $\bf B$ (three components), $p$, $\rho$ and the temperature $T$. An energy equation, or sometimes termed as the heat transport equation, is required to have a closed system. Here we adopt the approach of \citet{low1975} in which the energy equation is not solved directly. Once we have obtained a solution we can calculate the corresponding energy balance that the system has to satisfy in order to keep a thermal equilibrium, but this is not the main goal of the present study.

In 2D the equations can be written in terms of the flux function and the force balance leads to a Grad-Shafranov equation, this is the procedure adopted in \citet{terradasetal2022}. The functional dependence of pressure and temperature 
on
the flux function determines how the plasma is coupled to the magnetic field.  However, if we want to employ the equivalent approach in 3D the magnetic field needs to be written in terms of two EPs $\alpha({\bf r})$ and $\beta({\bf r})$ (e.g. \citealt{roberts1967,stern1970,stern1976}). In this case we have that 
\begin{align}
    {\bf B}=\nabla \alpha \times \nabla \beta \label{eqBEuler}.
\end{align}
Using vector identities it is easy to show that Eq.~(\ref{eq:divb}) is automatically satisfied.

The EPs are constant along the field lines of 
{\bf B} because ${\bf B} \cdot \nabla \alpha = 0$ and ${\bf B} \cdot \nabla \beta = 0$. Interestingly, this provides a method to compute $\alpha$ and $\beta$ in the domain when {\bf B} is known  \citep[see][and Sect.~\ref{sect:shear}]{stern1970}. This is achieved, for example, by
fixing the values of the EPs on the part of the boundary of positive polarity and transporting them into the domain along the lines. In this case the problem is linear.

From Eq.~(\ref{eqBEuler}) the magnetic field components in terms of the EPs read
\begin{align}
 B_x (x,y,z)&=\partial_y\alpha(x,y,z)\, \partial_z \beta(x,y,z)-\partial_z\alpha(x,y,z)\, \partial_y \beta(x,y,z),\label{eqBcompsBx}\\
 B_y(x,y,z) &=\partial_z\alpha(x,y,z)\, \partial_x \beta(x,y,z)-\partial_x\alpha(x,y,z)\, \partial_z \beta(x,y,z),\label{eqBcompsBy}\\
 B_z(x,y,z) &=\partial_x\alpha(x,y,z)\, \partial_y \beta(x,y,z)-\partial_y\alpha(x,y,z)\, \partial_x \beta(x,y,z). \label{eqBcompsBz}
\end{align}
These equations indicate that even in the situation of a known magnetic field, the calculation of the EPs is not trivial due to the products of partial derivatives.  According to Eqs.~(\ref{eqBcompsBx})-(\ref{eqBcompsBz})  each component of the magnetic field only depends on the derivatives of the EPs in the perpendicular direction to that component. Different methods to calculate the EPs are discussed in Sects.~\ref{sec:potentialfield} and \ref{sect:shear}.

The current density is
\begin{align}
    {\bf J}=\nabla \times (\nabla \alpha \times \nabla \beta), \label{eqJEuler}
\end{align}
and the corresponding components in Cartesian coordinates
contain partial derivatives of second order at most but each component contains up to 8 different terms. The complexity of the system has significantly increased with respect to the 2D case, described in terms of the flux function  or the vector potential.

It can be shown that in 3D the condition of force balance using the EPs reduces to the following coupled partial differential equations (see for example \citealt{birnschindler1981}, \citealt{schindler2006}, \citealt{neukirch2015} and references therein)
\begin{align}
    \nabla \alpha \cdot \nabla\times(\nabla \alpha \times \nabla \beta)&=-\mu_0\, \partial_{\beta} p, \label{eqFacrossa} \\
    \nabla \beta \cdot \nabla\times(\nabla \alpha \times \nabla \beta)&=\mu_0\,\partial_{\alpha} p,\label{eqFacrossb}\\
    \partial_z{p}&=-\rho g ,\label{eq:hydrop}
\end{align}
where $p(\alpha,\beta,z)$ is the gas pressure, $\rho(\alpha,\beta,z)$ is the plasma density, and we have assumed that the gravitational force is constant and pointing in the negative $z-$direction.
Both plasma pressure and density may generally depend on the two EPs and the gravitational potential, which in our case is identical to the $z-$coordinate up to a constant factor. We emphasise that the partial derivatives of the pressure in Eqs.~(\ref{eqFacrossa}) -- (\ref{eq:hydrop}) are to be taken under the condition that the other variables on which the pressure depends are kept constant. In particular the partial $z$-derivative in Eq.~(\ref{eq:hydrop}) is taken with the EPs being kept constant, i.e.~it is a derivative taken along field lines.
Using the ideal gas law, Eq.~(\ref{eq:ideal}), the most general solution to Eq.~(\ref{eq:hydrop}) is
\begin{align}
p(\alpha,\beta,z)=p_0(\alpha,\beta)\,e^{\textstyle -\int_0^z \frac{\bar{\mu} g }{\mathcal{R} T(\alpha,\beta,z')}\,dz'},\label{eqpressalongB}
\end{align}
where $T(\alpha,\beta,z)$ is the temperature profile that can depend on the $z$ coordinate as well. We define a reference pressure scale height as $H=\mathcal{R} T_0/\bar{\mu} g$, being $T_0$ a normalisation temperature normally taken as the coronal temperature. It is convenient to remark \citep[see also][]{low1975} that in principle $p_0(\alpha,\beta)$ and $T(\alpha,\beta,z)$ could be multivalued along the same field line. Here we adopt the simplest case where the same functional forms of $p_0(\alpha,\beta)$ and $T(\alpha,\beta,z)$ apply to all regions of
space. In this situation two points at the same height on any  magnetic field line have the same pressure and  temperature.

If gravity is neglected $p(\alpha,\beta)=p_0(\alpha,\beta)$, meaning that pressure is constant along magnetic field lines but can change from line to line.
Density is calculated from the ideal gas law using the known profiles for $p(\alpha,\beta,z)$ and $T(\alpha,\beta,z)$.
Equation~(\ref{eqpressalongB}) imposes a balance between the force due to the gas pressure gradient and the gravity force along the magnetic fields lines, while  Eqs.~(\ref{eqFacrossa}) and (\ref{eqFacrossb}) represent the condition of force balance perpendicular to the magnetic field. The two coupled PDEs are nonlinear and each equation contains up to 21 different terms plus the term due to gas pressure,
constituting a rather complicated system of coupled equations to solve. The equations are written in divergence form when the problem is solved numerically, see Appendix A for further details.

Equations~(\ref{eqFacrossa})-(\ref{eqFacrossb}) must be complemented with appropriate boundary conditions (BCs) at the limits of the domain. Here we consider an hexahedron, in particular a rectangular cuboid. The spatial size of the cuboid is $x_{min}\le x \le x_{max}$, $y_{min}\le y \le y_{max}$, and $z_{min}\le z \le z_{max}$ (we take $z_{min}=0$ in this work). Boundary conditions need to be imposed at the six sides of the cuboid to 
 be able to solve
the PDEs. The question that arises is which are the natural conditions that must be satisfied by the equilibrium equations in the variables $\alpha$ and $\beta$. The answer is found using  Grad's functional \citep{grad1964}.  It can be shown  that the first variation of the functional  provides information about the nature of the BCs \citep[see][for a detailed derivation]{zwingmann1987,schindler2006}.  There are two possibilities. The first case corresponds  Dirichlet conditions on $\alpha$ and $\beta$, meaning that the EPs are prescribed on the boundaries and they are not allowed to change. Physically, this boundary condition forces the location where a ﬁeld line
with labels $\alpha$ and $\beta$ cuts through the boundary, fixing the footpoints of the field lines. For several reasons that will become clear later, we chose the potential solutions (the current free situation) as the Dirichlet boundary conditions on our domain. The second case corresponds to homogeneous Neumann conditions. It is not difficult to show \citep[see][]{schindler2006} that these conditions  mean that the tangential component of the magnetic field on the boundaries is zero, i.e., the magnetic field is strictly normal to the boundary.

Since in our case we intend to reproduce the properties of a 3D AR, we prefer to focus on Dirichlet conditions and not forcing the magnetic field to be perpendicular to the faces of the box which seems to be rather artificial when applied to a curved magnetic field as that of a bipolar region. Neumann conditions are typically applied to the upper boundary when one considers the problem of 
magnetic configurations with different magnetic topology
inside the spatial domain \citep[see for example][]{zwingmann1987,plattneukirch1994}, but this is out of the scope of the present work.

\section{General results for static ideal plasmas 
in
equilibrium}

It is appropriate to recall some known results in MHD \citep[see for example][]{roberts1967} that are be useful to interpret some of the general properties of MHS equilibria like the ones we obtain later from our purely numerical calculations
\citep[see also][]{aly1989}. 

We begin by introducing the gas pressure and magnetic tensors
\begin{align}
    {\bf T_p}&={\bf I}p, \\
    {\bf T_B}&={\bf I}\frac{B^2}{2\mu_0}-\frac{{\bf B}{\bf B}}{\mu_0},
\end{align}
where {\bf I} is the unit dyadic tensor. The static equation of motion given by Eq.~(\ref{eq:mhdstatic}) reads now
\begin{align}
    0=-\nabla\cdot\left({\bf T_p}+{\bf T_B}\right)+\rho {\bf g}.\label{eq:mgdstatictensorV}
\end{align}
If we consider a general volume $V$ with surface $S$, integrating Eq.~(\ref{eq:mgdstatictensorV}) over this volume and using Gauss's theorem we obtain
\begin{align}
    0=-\int_S \left({\bf T_p}+{\bf T_B}\right)\cdot d{\bf S}+\int_V \rho {\bf g}\, dV, \label{eq:mgdstatictensorS}
\end{align}
where $d{\bf S}= \mathbf{n}\, dS$ 
is the directed surface element and  
$\mathbf{n}$
the outward normal vector 
of
the surface. The gravitational force is left as a body force here, but for self-gravitating system (different to our case) it would be more convenient to write it as a gravitational stress tensor similarly to the gas pressure and magnetic field.

The terms that appear in Eq.~(\ref{eq:mgdstatictensorS}) involve surface integrals and are inevitably related to the BCs that are imposed in the spatial domain. It is therefore convenient to understand clearly the role of the BCs in the present problem.
The basic type of BCs used in this work are Dirichlet conditions, i.e., the EPs $\alpha(x,y,z)$ and $\beta(x,y,z)$ are imposed on the 
side boundaries 
of the system. This means, according to Eqs.~(\ref{eqBcompsBx})-(\ref{eqBcompsBz}), that $B_n={\bf n}\cdot {\bf B}$, i.e., the normal component of the magnetic field is prescribed by these BCs. However, the magnetic field component lying in
the plane of each boundary, i.e., $\bf B_t$, depends on the behaviour in the interior points of the domain and are not enforced. They adjust according to the solution achieved inside the domain. These magnetic field components modify the inclination of the magnetic field at the boundary and change the values of the magnetic stress tensor $\bf T_B$.

We return to Eq.~(\ref{eq:mgdstatictensorS}) which indicates that to have equilibrium there must be a  balance between forces on the surface of the volume, conditioned by the BCs,  and the gravitational volume force. When gas pressure and gravity are neglected the total magnetic stress on the surface of the volume must be zero. This does not necessarily mean that the magnetic stress is zero at all the points on the surface (in this case it is known that ${\bf B}=0$), it is the integrated value that is zero.

Another interesting general result, closely related to the conservation of energy, is the virial theorem which is useful, among other things, to check the validity of our numerical calculations in the following sections and to extract conclusions about the behaviour of the system. In the static case the virial theorem including gravity reads \citep[e.g.][]{chandra61,mckeezweibel1992,kulsrud2004}
\begin{align}
    0=3(\gamma-1)E_p+E_B-\int_S {\bf r}\cdot \left({\bf T_p}+{\bf T_B}\right)\cdot d{\bf S}-E_g.\label{eq:virialn}
\end{align}
where ${\bf r}$ is a radius vector relative to an origin chosen to be inside the volume $V$ and $\gamma$ is typically taken to be $5/3$. In the previous equation the total internal energy is defined as
\begin{align}
   E_p=\int_V \frac{p}{\gamma-1} dV, \label{eq:internal}
\end{align}
the total magnetic energy is given by
\begin{align}
   E_B=\int_V \frac{B^2}{2 \mu_0} dV, \label{eq:magnetic}
\end{align}
while the gravitational energy is
\begin{align}
   E_g=-\int_V {\bf r}\cdot \rho {\bf g}\, dV. \label{eq:gavit}
\end{align}
It is important to realise that due to the Dirichlet conditions that we use, the total magnetic energy in the system changes when, for example, gas pressure is modified, meaning that $E_B$ is not maintained constant even in the case when the same values of $\alpha$ and $\beta$ are used for the different equilibria. This is due to the changes in ${\bf T_B}$ which has two contributions: one that is fixed in the present work due to Dirichlet boundary conditions, and another term that changes to obtain an equilibrium under the presence of gas pressure and gravity.

The first two terms in Eq.~(\ref{eq:virialn}) are always positive but the last two terms  might be negative and capable to provide the required balance in the equation. The third term is due to the surface pressure stress and magnetic stress,  while the last term is due to the integrated gravity on the volume.

\section{The potential magnetic field in 3D}\label{sec:potentialfield}

To find equilibrium solutions we have shown that we need to provide the values of $\alpha$ and $\beta$ at the boundaries (Dirichlet BCs). The EPs at the boundaries determine the global geometry of the magnetic field inside the box and they must be chosen to represent the specific magnetic structure that is of interest, in our case closed magnetic field lines representative of ARs. As a starting point we use EPs that have an analytical expression, this is typically the case 
for
potential magnetic fields with some symmetry axis. 

\subsection{The magnetic bipole}\label{subsect:bipole}

One of the elementary magnetic arrangements that one can imagine in 3D relies on the superposition of fictitious monopoles that decay with distance as $1/r^2$,
\begin{align}
 {\bf B}=a_0 \, \frac{\hat {r}}{r^2}, \label{eqBmonopole}
\end{align}
where $a_0$ is a constant. We start by superposing two magnetic monopoles of opposite polarity situated below our reference plane at $z=0$ at a depth $z=-z_0$ and separated a distance $d$ from the origin. The magnetic field components of this structure in Cartesian coordinates are
\begin{align}
 B_x(x,y,z) /B_0&=L^2 \frac{x-d}{\big((x-d)^2+y^2+(z+z_0)^2\big)^{3/2}}\nonumber \\&-\frac{x+d}{\big((x+d)^2+y^2+(z+z_0)^2\big)^{3/2}},\label{eqBcompsbipole1}\\
 B_y(x,y,z)/B_0 &=L^2 \frac{y}{\big((x-d)^2+y^2+(z+z_0)^2\big)^{3/2}}\nonumber \\&-\frac{y}{\big((x+d)^2+y^2+(z+z_0)^2\big)^{3/2}},\label{eqBcompsbipole2}\\
 B_z(x,y,z)/B_0&=L^2 \frac{z+z_0}{\big((x-d)^2+y^2+(z+z_0)^2\big)^{3/2}}\nonumber \\&-\frac{z+z_0}{\big((x+d)^2+y^2+(z+z_0)^2\big)^{3/2}},\label{eqBcompsbipole3}
\end{align}
where $L$ is a distance used for normalisation purposes that is taken to be equal to the pressure scale height $H$ previously defined.  
This magnetic field is current-free and three-dimensional but it still has axial symmetry. This configuration has been used in the past by, for example, \citet{semel1988} and \citet{cupermanetal1989}. 

\begin{figure}[h]
\center{
\includegraphics[width=10cm]{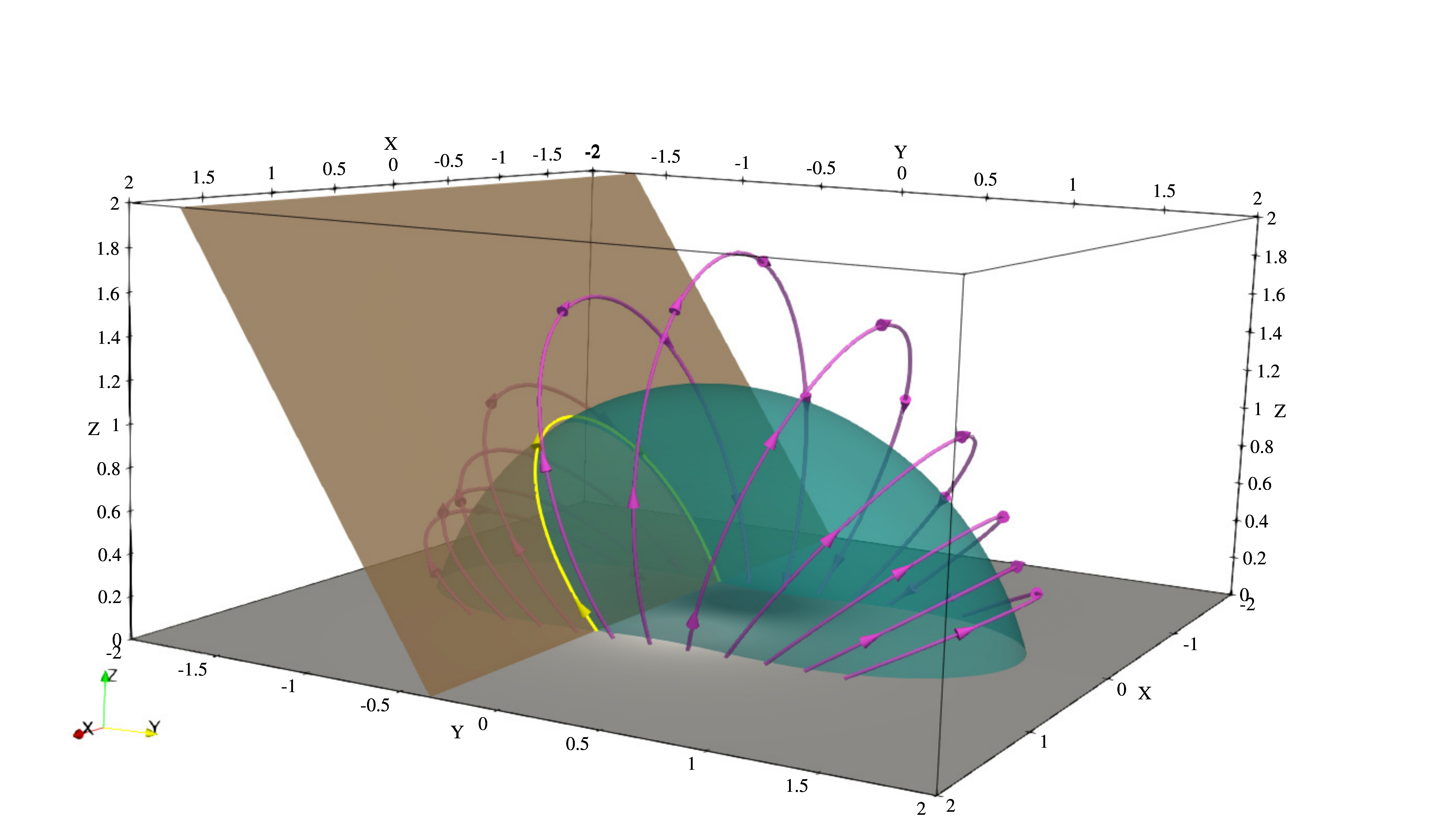}}
\caption{\small Potential magnetic field lines (pink curves with arrows) and two magnetic surfaces associated to the EPs $\alpha$ (green color) and $\beta$ (brown color) for a bipole with $d=0.25$ and $z_0=0.5$ (all the lengths are hereafter normalised to $H$). The intersection of the two surfaces, see the yellow curve, represents a particular magnetic field line. The distribution of the vertical component of the magnetic field at $z=0$ is represented in gray colors.} \label{fig:eulerpots}
\end{figure}

The next step is to describe the previous magnetic field in terms of the EPs, given by Eqs.~(\ref{eqBcompsBx})-(\ref{eqBcompsBz}). As explained in Sect.~\ref{sec:mhs} these equations are nonlinear and in general difficult to solve for $\alpha$ and $\beta$ even if  $B_x$, $B_y$, $B_z$ are known functions like in the present case. However, the symmetry of the magnetic bipole
(rotational symmetry with respect to the line that connects the two monopoles) allows us to derive the corresponding EPs. 
As a geometrically simpler example, we first consider 
an arcade that is invariant in the $y-$direction, e.g. the arrangement investigated by \citet{zwingmann1987}. In the case without shear the second EP is $\beta=y$ while $\alpha$ is just the flux function (or vice versa). Isocontours of $\beta$ are vertical planes that are labeled according to the value of $y$. Applying this geometrical property to the  rotationally symmetrical bipole, we simply have that $\beta=\theta$, with $\theta$ being the angle with respect to the line that connects the two monopoles. Therefore, in Cartesian coordinates we have that
\begin{align}
 \beta(y,z) &=\arctan{\frac{y}{z+z_0}}.\label{eqEulerpotbipolebet}
\end{align}
Since $\partial_x \beta=0$ it is easy to calculate the EP $\alpha$ by direct integration of Eqs.~(\ref{eqBcompsBx})-(\ref{eqBcompsBz}),
\begin{align}
 \alpha(x,y,z)/\alpha_0&=\frac{x-d}{\sqrt{(x-d)^2+y^2+(z+z_0)^2}}\nonumber \\&-\frac{x+d}{\sqrt{(x+d)^2+y^2+(z+z_0)^2}}.\label{eqEulerpotbipolealpha}
\end{align}
An example of the spatial distribution of $\alpha$ and $\beta$ in 3D based on the previous expressions is shown in Fig.~\ref{fig:eulerpots}. The intersection of different isosurfaces of constant $\alpha$ and $\beta$ coincide with the magnetic field lines. Hence, each magnetic field line is characterised by a pair of values $(\alpha,\beta)$.

It is not difficult to check that the EPs given by Eqs.~(\ref{eqEulerpotbipolebet}) and (\ref{eqEulerpotbipolealpha}) satisfy, according to Eq.~(\ref{eqJEuler}), that $J_x=J_y=J_z=0$, i.e., the magnetic field is potential and Eqs.~(\ref{eqFacrossa}) and Eqs.~(\ref{eqFacrossb}) are fulfilled since the gas pressure has no effect on the equilibrium configuration in 
this case.

 Since the potential solution has rotational symmetry it is convenient to use cylindrical instead of Cartesian coordinates. The solution given by Eqs.~(\ref{eqEulerpotbipolebet})-(\ref{eqEulerpotbipolealpha}) reads in cylindrical coordinates
\begin{align}
 \beta(\theta) &=\theta,\label{eqEulerpotbipolebetcyli}
\end{align}
as indicated earlier, while
\begin{align}
 \alpha(r,z)/\alpha_0&=\frac{z-d}{\sqrt{r^2+(z-d)^2}}-\frac{z+d}{\sqrt{r^2+(z+d)^2}},\label{eqEulerpotbipolealphacyli}
\end{align}
being independent of $\theta$. The $z-$coordinate is pointing now along the symmetry axis in this new coordinate system (the $x-$direction in the Cartesian coordinates).

The EPs are 
not
unique since a 
different
gauge leads to the same magnetic field, but it is convenient to chose the most simple geometrical surfaces for the EPs. For example, we have seen that for the symmetric bipolar magnetic field $\beta$ can be represented by planes, and this is our preferred option, see Fig.~\ref{fig:eulerpots}.
\subsection{The general analytic solution for the case with axial symmetry}\label{analyticcalc}
 If we focus on the 
 case in which the pressure gradient either vanishes or is in
 hydrostatic equilibrium with the gravitational force,
 the potential structure in cylindrical coordinates is a solution to the following equations \citep[see also][]{kaisersalat1997}
 \begin{align}
    \partial^2_{rr}\alpha-\frac{1}{r}\partial_r \alpha+\partial^2_{zz}\alpha&=0,\label{eqnoLaplacecyl}\\
    \partial_\theta\left(\frac{1}{2\mu_0}\frac{1}{r^2}\left[(\partial_r \alpha)^2+(\partial_z \alpha)^2\right]\right)&=0.
    \label{eqpress2Dtot}
\end{align}
Assuming that $\alpha$ is independent of $\theta$ the second equation is automatically satisfied and we only need to solve the first equation. 
The simple bipole model given by Eq.~(\ref{eqEulerpotbipolealphacyli}) satisfies this equation. However, it is useful to generalise the method of finding the potential solution for any magnetic spatial distribution on the axis, $\alpha(0,z)$, and this is the purpose of the present section. As far as we know the following derivation has not been reported in the literature.

In order to avoid undesirable boundary effects we wish to solve Eq.~(\ref{eqnoLaplacecyl}) by imposing that $\alpha=0$ for $r\rightarrow \infty$. We use the method of separation of variables and assume that
\begin{align}
    \alpha(r,z)=f(r)\,h(z),
\end{align}
which leads to 
\begin{align}
    \frac{f''(r)}{f(r)}-\frac{1}{r}\frac{f'(r)}{f(r)}=-\frac{h''(z)}{h(z)}=k^2,\label{eqsepvar}
\end{align}
being $k$ the separation parameter and the derivatives are with respect to the arguments of the functions. We obtain two separate ODEs that share the parameter $k$. The easiest ODE to solve corresponds to the $z-$dependence which reads 
\begin{align}
    h''+k^2 h=0.
\end{align}
The solution to this equation is a superposition of a sine and a cosine function, 
\begin{align}
    h(z)=A \cos\left(k z\right)+B \sin\left( k z\right).
\end{align}
being $A$ and $B$ some constants that need to be determined. The separation parameter $k$ appears in the argument of the functions.

Returning to Eq.~(\ref{eqsepvar}), the ODE for the radial part is
\begin{align}
    f''-\frac{1}{r}f'-k^2 f=0.\label{besselmod}
\end{align}
To solve this equation we assume that $f(r)=r \mathcal{F}(kr)$ which is inserted into  Eq.~(\ref{besselmod}). This process leads to a modified Bessel equation for $\mathcal{F}(kr)$, meaning that the general solution to Eq.~(\ref{besselmod}) is of the form
\begin{align}
    f(r)=r\Big[C\, I_1(kr)+D\, K_1 (kr)\Big].
\end{align}
where $I_1(kr)$ and $K_1(kr)$ are the modified Bessel functions of order one while $C$ and $D$ are 
constants that need to be determined according to the BCs.

For $r \rightarrow \infty$ we have that $r I_1(kr)\rightarrow \infty$ and therefore we have to impose $C=0$ to avoid this behaviour. On the contrary, $r K_1(k r)\sim \sqrt{r} e^{-kr}$ for large $r$ which means that this solution goes to zero at infinity. For $r \rightarrow 0$, $r K_1(k r)\simeq 1/k$ and has the correct 
behaviour 
at the origin (finite and different from zero in general).

The formal solution to our problem taking into account the $z$ and $r$ dependencies is the following superposition
\begin{align}
    \alpha(r,z)=\int_0^\infty \big[A(k) \cos(k z)+B(k) \sin(k z)\big]\, r\, K_1(k r) \, dk.\label{eqsolpot}
\end{align}
The Fourier coefficients $A(k)$ and $B(k)$ are determined from the chosen profile for $\alpha$ at $r=0$ (i.e., on the axis), which  according to the previous expression is
\begin{align}
    \alpha(0,z)=\int_0^\infty \big[A(k) \cos(k z)+B(k) \sin(k z)\big]\frac{1}{k} \, dk,
\end{align}
where the value at the origin of $r K_1(k r)$ has been taken into account. Using the known result from Fourier analysis the coefficients read
\begin{align}
\frac{A(k)}{k} &=\frac{1}{\pi} \int_{-\infty}^{\infty}
\alpha(0,s) \cos \left(k s\right) ds, \label{eq:Acoef}\\
\frac{B(k)}{k} &=\frac{1}{\pi} \int_{-\infty}^{\infty}
\alpha(0,s) \sin \left(k s\right) ds. \label{eq:Bcoef}
\end{align}
If $\alpha(0,z)$ is a symmetric function with respect to $z=0$ then $B(k)=0$, while if it is anti-symmetric $A(k)=0$. 

The main result of this section is that we have found a general expression, Eq.~(\ref{eqsolpot}), that allows us to calculate the potential solution in the $r-z$ plane given any  profile of $\alpha$ on the axis (i.e., $\alpha(0,z)$). This solution is used later as BC to construct MHS equilibria in 3D under the presence of gas pressure and gravity. It is also adopted as the starting point to include magnetic shear in the structure. A simple test of the validity of the previous expressions is achieved by reproducing the  double  monopole potential solution given by Eq.~(\ref{eqEulerpotbipolealphacyli}).

A similar approach can be applied to the case with 
the effect of the
gas pressure 
included,
as long as the derivative of this magnitude with $\alpha$ is proportional to $\alpha$, i.e., the linear case \citep[see the analysis performed in][]{atanasiuetal2004}. When gravity is 
also
present,
no analytic solutions are available, as far as we know.

The relevance of the procedure used before to construct solution becomes clear when 
we consider an example of a structure that lacks symmetry along the axis, i.e. that is not symmetric around $z=0$ (in cylindrical coordinates). The first step is to chose the particular form of $\alpha(0,z)$. In order to simplify the calculations we chose a profile that leads to analytical results for the Fourier coefficients. Our deliberate choice is a Gaussian
\begin{align}
    \alpha(0,z)=C\,e^{-\left(\frac{z-z_s}{w}\right)^2},
\end{align}
situated at $z_s$ (not necessarily equal to zero) and with a characteristic width $w$ being $C$ a constant.  Substituting this expression into Eqs.~(\ref{eq:Acoef}) and (\ref{eq:Bcoef}) and performing the integrals \citep[see][]{gradshteyn2007} yields 
\begin{align}
A(k)&=C \sqrt{\frac{1}{\pi}} w k \,e^{-\frac{w^2}{4} k^2}\cos(z_s k),\\
B(k)&=C \sqrt{\frac{1}{\pi}} w k \,e^{-\frac{w^2}{4} k^2}\sin(z_s k).
\end{align}
The solution $\alpha(r,z)$ is obtained by substituting the previous coefficients in Eq.~(\ref{eqsolpot}) and integrating with respect to $k$. This integral over the semi-infinite range does not have an analytical solution and must be calculated numerically. We have used \textsc{Mathematica} to compute the integral.

\begin{figure}[h]
\center
\includegraphics[width=12cm]{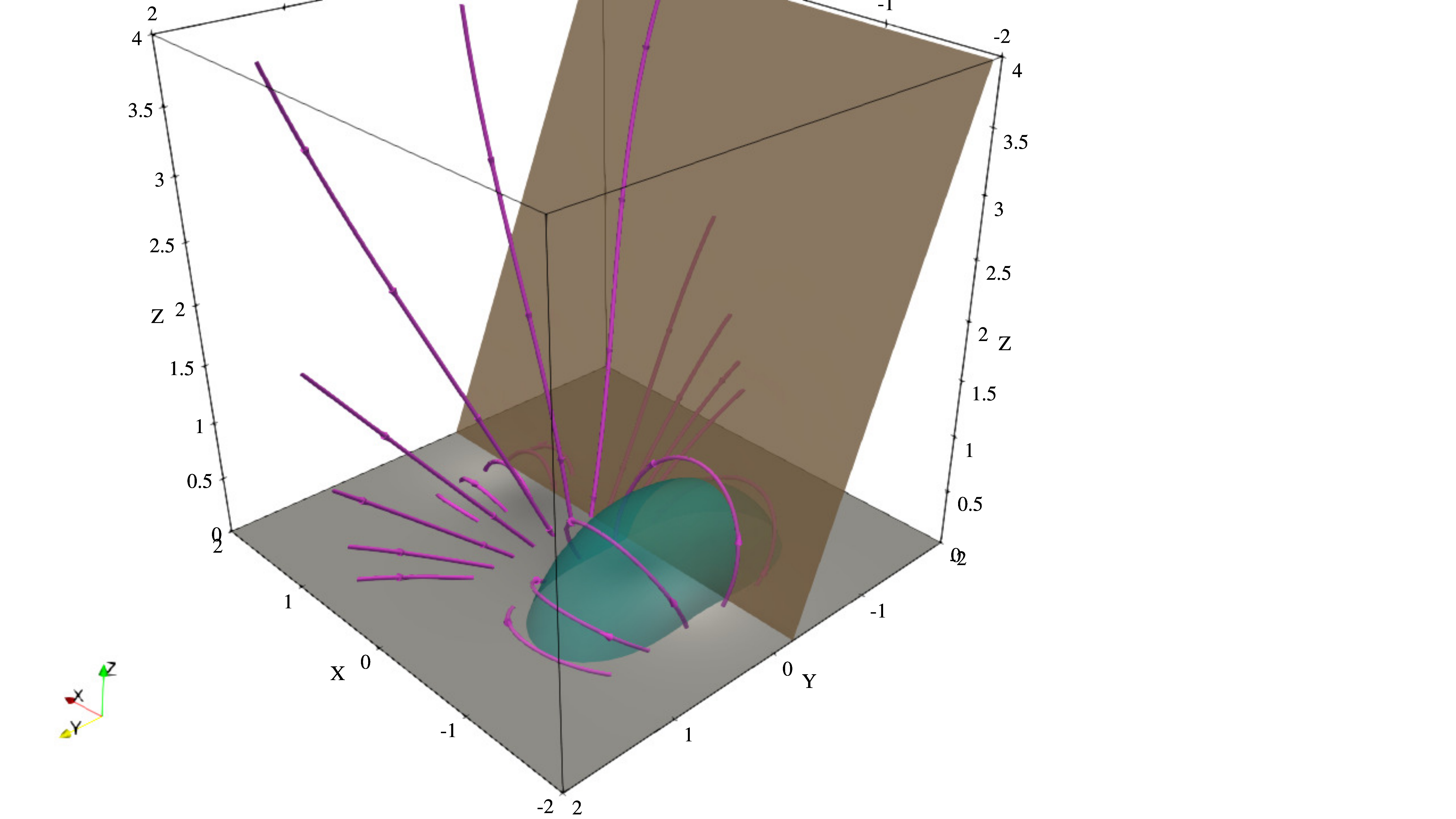}
\caption{\small Potential magnetic field lines and isosurfaces  calculated numerically for a non-symmetric case along the $x-$direction. In this example, the following parameters have been used, $w_1=w_2=1/4$, $z_1=-3/4$, $z_2=1$ and $C_1=1$, $C_2=-0.5$. } \label{fig:eulerpotnosymm}
\end{figure}

Since Eq.~(\ref{eqnoLaplacecyl}) is a linear PDE, a superposition of, for example, several Gaussian profiles at distinct locations and with different amplitudes is readily constructed using the previous expressions. Once the profile $\alpha(r,z)$ is known a change from cylindrical to Cartesian coordinates is needed and taking into account that our reference level is located at $z_{min}$  the symmetry axis in cylindrical coordinates is situated at a depth $z_d$ below $z_{min}$. In Fig.~\ref{fig:eulerpotnosymm} we show an example of the superposition of two Gaussians with a certain values of the parameters. In this case the two constants $C_1$ and $C_2$ have opposite signs. The constructed structure has a central region where the magnetic field is open, as we see on the top face of the box, and could represent a CH that is surrounded by two AR with closed field lines. The configuration still has rotational symmetry, because surfaces of constant $\beta$ are still planes.

\begin{figure}[h]
\includegraphics[width=12.5cm]{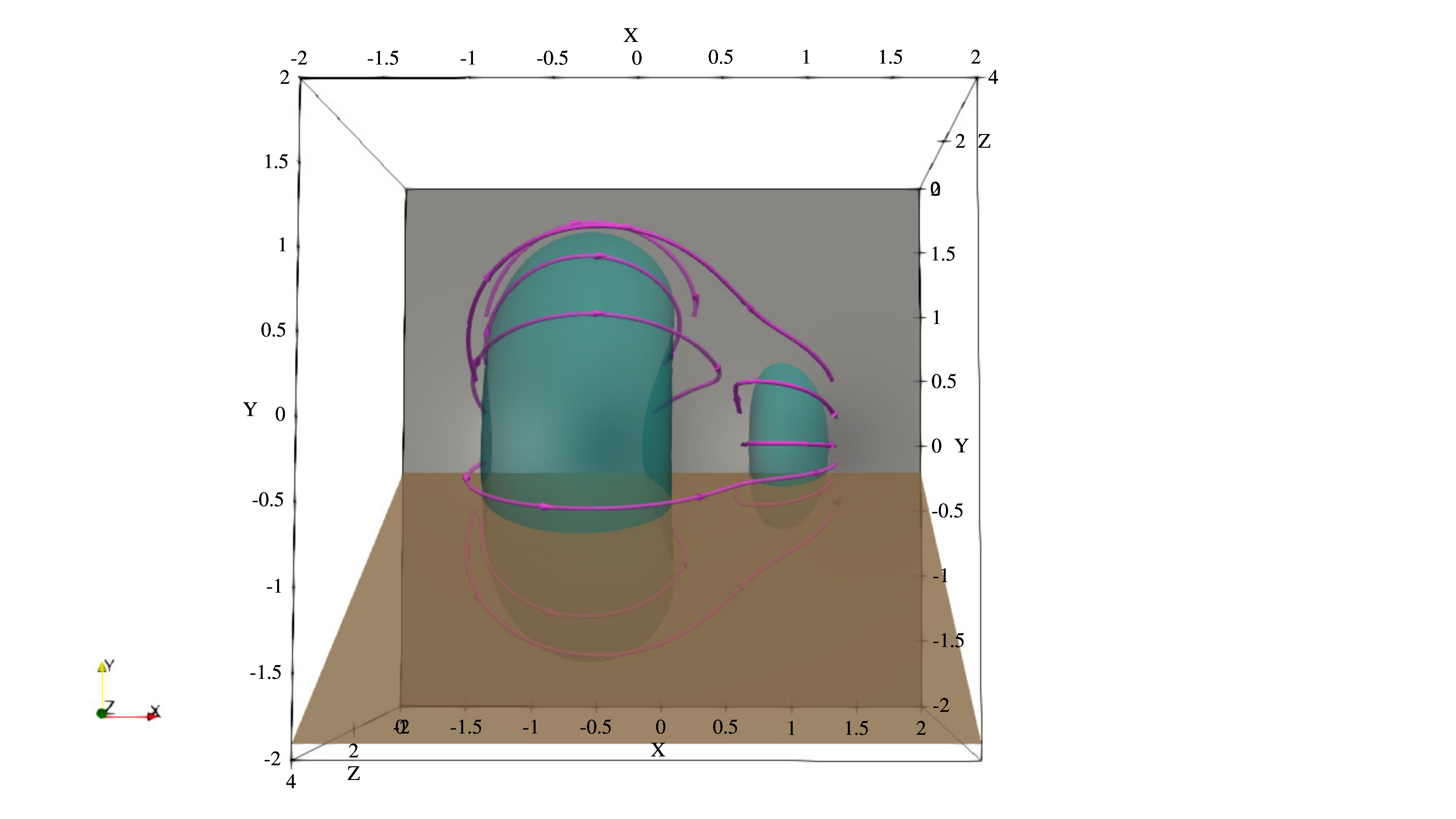}
\caption{\small Potential magnetic field lines and isosurfaces that include $X-$points. A visible critical point is located around $x=0.5$ and $y=0.5$ but due to rotational symmetry it is repeated along a circular path, i.e., it is in reality a null line. The same parameters as in Fig.~\ref{fig:eulerpotnosymm} have been used except that $C_1$ and $C_2$ have the same sign now.} \label{fig:eulerpotxpoint}
\end{figure}

Interestingly, if we assume that the factors $C_1$ and $C_2$ have the same sign, the opposite of the situation shown in Fig.~\ref{fig:eulerpotnosymm}, the topology of the magnetic field changes and it may happen that at some points the magnetic field becomes zero, i.e., there are  
magnetic null points. According to the definition of the magnetic field in terms of the EPs and due to the symmetry in the azimuthal direction assumed in this section ($\beta=\theta$) the only possibility to satisfy ${\bf B}=0$ is that $\nabla \alpha=0$. Since $\alpha$ depends on $r$ and $z$ only, the last condition means that at 
a magnetic null point,
which is precisely a critical point in a mathematical sense, $\alpha(r,z)$ has either a local minimum, a local maximum or a saddle point. 
Minima or maxima of $\alpha$ correspond to $O$-type magnetic nulls,
whereas saddle points correspond to $X$-type null points.
An example that includes an $X-$type magnetic field structure is shown in Fig.~\ref{fig:eulerpotxpoint}. 
In reality, this is actually 
a null line (a curve where the magnetic field vanishes) due to the rotational symmetry. 
This is therefore essentially an $X-$point in a 2D sense, since the characteristics of a truly 3D null point are not present in this example, there is no spine line or fan plane \citep[see for example][]{parnelletal1996,priestforbes2007}. 
The connectivity
of the magnetic 
field lines
changes when moving across
the separatrix surfaces which intersect at
the null line. 
It is known that, contrary to isolated 3D null points, null lines can be locally described using EPs \citep[see][]{hesseschindler1988}, as in the present case. However, it is also known that magnetic null lines are structurally unstable, i.e. any arbitrarily small additional magnetic field will either generate a magnetic field without nulls or one with isolated nulls \citep[e.g.][]{schindleretal1988}.

\section{The non-potential magnetic field in 3D: incorporating the effect of shear}\label{sect:shear}

The general potential solutions presented in the previous section have rotational symmetry with respect the axis underneath $z_{min}$ where we prescribe the form of $\alpha$. These solutions do not explicitly include shear and hence no field-aligned component of the current density. 
In the present section we discuss different methods to incorporate magnetic shear to the magnetic structure yielding to non-potential solutions. The problem is not straightforward since we need to work with the EPs instead of the magnetic field components. There are several possible approaches to attack this problem.

The first method is based on the use of known analytical expressions for sheared bipolar magnetic configurations. For example, \citet{cupermanetal1989} give the three magnetic field components in terms of a parameter that measures the amount of shear in the structure, this parameter is called here $a$ ($a=0$ corresponds to the potential solution given by Eqs.~(\ref{eqBcompsbipole1})-(\ref{eqBcompsbipole3})). Even with  the known expressions for the magnetic field components the calculation of the corresponding EPs when $a$ is different from zero is not an easy task due to the nonlinear character of the equations. One possibility is to use an asymptotic expansion, as in for example \citet{birnschindler1981} \citep[see the explanation for the technique in][]{schindler2006}, to obtain approximate expressions for the EPs. It turns out that this approach is rather involved and we have not been able to find analytical or semi-analytical solutions.  Another option is to calculate $\alpha$ and $\beta$ numerically using the computation of the magnetic field lines, known in this case, as it was proposed by \citet{stern1970}. The idea is that if the EPs are known on a given surface, then tracking the intersection of any field line with this surface provides the values of the Euler parameters on the magnetic field lines and therefore on any point of the domain (as long as all the field lines intersect the reference surface). The foremost limitation of this method is that we must know first the magnetic field distribution in 3D in order to calculate the EPs, and this is not the typical situation. 

We propose here a method that we think it is more flexible that the previous approaches. It is based on transforming the Euler variables in the potential situation in such a way that they lead to magnetically sheared states. Again gas pressure and gravity are added once the magnetically force-free solution is obtained. The method is precisely based on directly shearing the EPs at $z=0$ and also at the other faces of the box but still keeping a rectangular shape for the domain. To do so we apply an affine coordinate shear transformation defined by $x'=x+s_y y$ and  $y'=s_x x+y$, where $s_x$ and $s_y$ are the shear factors in the $x$ and $y-$directions. This transformation produces the effect that we seek for on the magnetic field as it is demonstrated in the following. For example, the vertical component of the magnetic field at $z=0$ is according to Eq.~(\ref{eqBcompsBz})
\begin{align}
 B_{z0}(x,y,0) &=\partial_x\alpha(x,y,0)\, \partial_y \beta(x,y,0)\nonumber \\&-\partial_y\alpha(x,y,0)\, \partial_x \beta(x,y,0), \label{eqBcompsBzsh}
\end{align}
and the shear transformation changes $B_{z0}$ to $B_{z1}$
\begin{align}
 B_{z1}(x,y,0) &=\partial_x\alpha(x',y',0)\, \partial_y \beta(x',y',0)\nonumber\\&-\partial_y\alpha(x',y',0)\, \partial_x \beta(x',y',0). \label{eqBcompsBzsh1}
\end{align}
Using the chain rule for multivariable functions and the coordinate transformation it is not difficult to find that the previous expression reduces to
\begin{align}
 B_{z1}(x,y,0) &=B_{z0}(x',y',0) \left(1-s_x s_y\right), \label{eqBcompsBzsht}
\end{align}
meaning that the final magnetic field is the initial magnetic profile but sheared, due to the presence of the 
%
primed coordinates
in the argument of $B_{z0}$, times a factor that is proportional to the shear parameters in each direction. The same applies to the magnetic field components perpendicular to the rest of the faces of the box.


The previous idea of applying a deformation on the EPs is developed further. We define another transformation 
\begin{align}
x'&=x+g(y),\\
y'&=y+f(x),
\end{align}
which can be shown that leads to
\begin{align}
 B_{z1}(x,y,0) &=B_{z0}(x',y',0) \left(1-\frac {df}{dx} \frac{dg}{dy}\right). \label{eqBcompsBzgen}
\end{align}
Now the functions $f(x)$ and $g(y)$ can be conveniently chosen to make the parenthesis of the previous equation to be independent of $x$ and $y$, ensuring that the transformation on the EPs produces exactly the expected shear on the vertical component of the magnetic field. The shear transformation previously discussed fits into this category as a particular case (where $f(x)=s_x x$, $g(y)=s_y y$) but there are other alternatives that allow us to incorporate families of changes on the magnetic field (that are not necessarily affine transformations). For example, we can start with a potential solution that is axially symmetric and trough a transformation achieve a non-symmetric solution. To this purpose we chose 
\begin{align}
g(y)&=0, \label{eq:sheardispl0} \\ f(x)&=a_0\,e ^{-\left(\frac{x-x_0}{w}\right)^2}.\label{eq:sheardispl}
\end{align}
This transformation makes the parenthesis of Eq.~(\ref{eqBcompsBzgen}) to be one and produces the effect of shifting in the $y-$direction an amount $a_0$ the magnetic field around $x=x_0$ within a typical distance $w$. This is a rather simple way of creating non-symmetric profiles for the EPs starting from  symmetric known solutions. 

Examples of the application of the two transformations  on the EPs explained above, that lead to particular sheared magnetic field configurations, are discussed in Sect.~\ref{sec:shear}. 

\section{The numerical method}\label{sec:method}

Once that we have an initial magnetic distribution uncoupled 
from
the plasma we aim at finding solutions to Eqs.~(\ref{eqFacrossa}) and~(\ref{eqFacrossb}). Unfortunately, analytical solutions to these equations are extraordinarily difficult to obtain, especially if the effect of gas pressure and gravity are included in the model. Ignoring gravity and gas pressure may lead to analytical solutions under some symmetry conditions and some example have been shown in Sect.~\ref{analyticcalc}. But in general numerical techniques are required to obtain solutions to the equilibrium equations. In the present work and in order to solve the PDEs with the corresponding boundary conditions we use the finite element method (FEM) \citep[for details see, e.g.][]{zienkiewiczetal2013,ganesantobiska2017}. 

We have used the FEM implemented in
\textsc{Mathematica} 
to solve linear and nonlinear PDEs. In general, we have found a good performance of the software, which uses parallelisation to speed up the computation time. The PDEs must be introduced into the code in divergence form (see Appendix A).  Apart from the BCs at the six faces of the cuboid an initial 
condition
for the solution is required by the numerical algorithm. This 
initial condition
has been chosen to be the potential solution calculated using the analytical expressions of Sect.~\ref{sec:potentialfield} and applying the procedure to include shear explained in Sect.~\ref{sect:shear}. The problems addressed in the present work do not require the use of continuation methods to calculate sequences of equilibria and detect the presence of bifurcation points \citep[see for example][]{neukirch1993a,neukirch1993b,romeouneukirch1999}. The simple strategy of slowly varying 
a parameter, calculate a solution and use it as the initial 
condition 
for the following step has been shown to be sufficient for the type of equilibria considered in this paper.

\begin{figure}[h]
\center
\includegraphics[width=7cm]{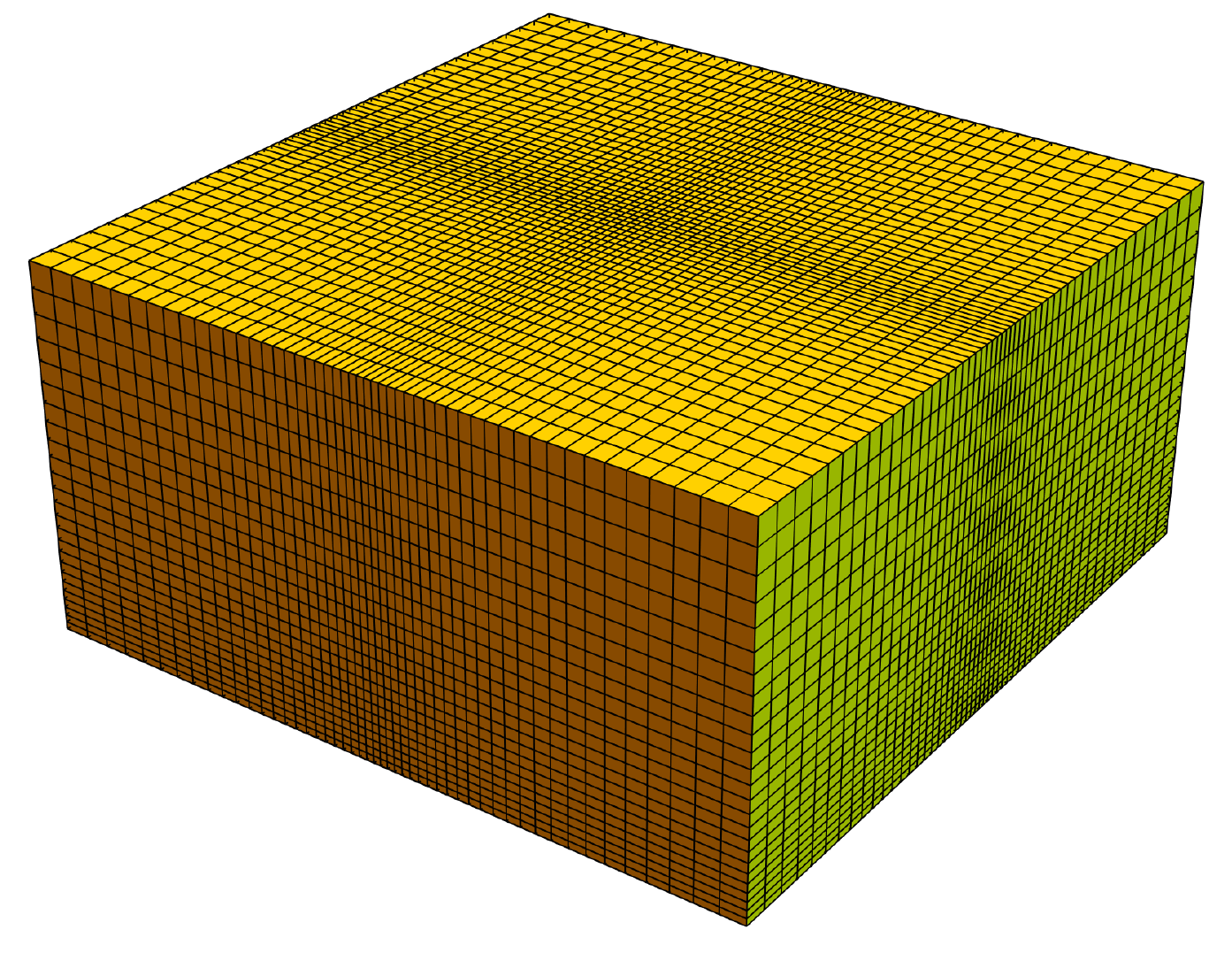}
\caption{\small Typical mesh structure used in the finite element calculations. The $x-$axis and $y-$axis point along  the orange and green faces of the box, while the $z-$axis is vertical. A non-uniform grid in each direction has been used to improve the accuracy of the calculations. The most refined region of the grid is located around $x=y=z=0$. In this example the grid has 62,500 elements.} \label{fig:mesh}
\end{figure}

In most of the calculations we have used a 
non-uniform numerical
mesh with refinement around the region where the strongest magnetic field is prescribed, i.e., around the center of the face $z=z_{min}$. An example of the mesh is shown in Fig.~\ref{fig:mesh}. In this configuration the mesh is basically the product of three non-uniform meshes, two of them refined around the center (in the $x$ and $y$ directions) and another refined around $z_{min}$.

Several
tests have been performed to validate the obtained numerical solutions. The first 
test is to consider the potential case (no gas pressure or gravity) and use the analytical potential solution given by Eqs.~(\ref{eqEulerpotbipolebet}) and (\ref{eqEulerpotbipolealpha}) as boundary conditions at the six faces of the cuboid and also as the initial 
condition
in the whole 3D box. The discrepancies between the numerically computed solution and the analytical results are very small and decrease as the number of mesh elements is increased. For non-potential calculations, using again the potential solution as BC and as the 
initial condition,
we have calculated the corresponding volumetric and surface integrals in Eq.~(\ref{eq:virialn}) for the numerically obtained solutions. The computations indicate that the numerical error in the expression of the virial theorem is typically around $2 \%$ and becomes smaller when the number of finite elements is increased. This is a clear indication that the calculated equilibria, under the presence of gas pressure and gravity are correct. 

The numerical difficulties found in the present problem are in some cases related to the fact that the 
the initial condition is not sufficiently close to the actual 
solution. 
Another difficulty is due to the potential solution imposed at the boundaries, especially at the upper boundary, which may lead to convergence issues because the system is forced to satisfy the BCs even under the presence of other forces such as the gas pressure gradient and the gravity force. This last issue is more relevant in the case when the plasma-$\beta$
exceeds the value of
one near the upper boundary.

\section{Equilibrium calculations: numerical results}\label{sect:numequil}

The inclusion of gas pressure and gravity changes the behaviour of the magnetic field and leads to a non-potential structure that needs to be computed. The functional dependencies of gas pressure and temperature
on 
the EPs allows us the calculation of a wide range of different equilibrium configurations. We begin with the most simple cases and then the complexity of the models is gradually increased.

Hereafter we assume that temperature does not depend explicitly on the coordinate $z$ (but this is not a constraint in the equations considered here). In this case Eq.~(\ref{eqpressalongB}) leads to the exponentially stratified atmosphere of the form
\begin{align}
p(\alpha,\beta,z)=p_0(\alpha,\beta)\,e^{-\frac{\bar{\mu} g }{\mathcal{R} T(\alpha,\beta)} z}.\label{eqpressalongBs}
\end{align}
The isothermal assumption along the magnetic field lines has some implications regarding the stability of the system that will be discussed in Sect.~\ref{sectinstability}. For the moment we concentrate on the analysis of different equilibrium configurations.

\subsection{Functional dependence 
on $\alpha$}

In order to obtain an equilibrium we have the freedom to prescribe the functional dependence of pressure and temperature 
on
$\alpha$ and $\beta$. For simplicity we focus on the situation that is independent of $\beta$. A useful functional dependence for gas pressure is 
\begin{align}
p_0(\alpha)=p_{\rm C}+(p_{\rm AR}-p_{\rm C})\left(\alpha/\alpha_{\rm ref}\right)^2,\label{eqp_0}
\end{align}
where $p_{\rm C}$ is the coronal pressure, $p_{\rm AR}$ the AR pressure and $\alpha_{\rm ref}$ a reference value used for normalisation purposes. We assume that $p_{\rm AR}>p_{\rm C}$ and since the dependence in Eq.~(\ref{eqp_0}) is with the square of $\alpha$ we ensure that gas pressure is positive everywhere ($\alpha$ is not necessarily a positive number). A possible choice for the temperature is, 
\begin{align}
T(\alpha)=T_{\rm C}+(T_{\rm AR}-T_{\rm C})\left(\alpha/\alpha_{\rm ref}\right)^2,
\end{align}
where $T_{\rm C}$ is the coronal pressure and  $T_{\rm AR}$ the temperature at the core of the AR (we assume that $T_{\rm AR}>T_{\rm C}$). Applying the ideal gas law at the center of the structure and at the reference coronal part it is not difficult to find the following relationship
\begin{align}
\frac{\rho_{\rm AR}}{\rho_{\rm C}}=\frac{p_{\rm AR}}{p_{\rm C}} \frac{T_{\rm C}}{T_{\rm AR}}.\label{eq:denscontrast}
\end{align}
The condition $p_{\rm AR}/p_{\rm C}> T_{\rm AR}/T_{\rm C}$  needs to be satisfied in order to represent an overdense region with respect to the environment. We typically impose that $T_{\rm AR}/T_{\rm C}=2$, i.e., the AR core is assumed to be twice as warm as the coronal environment, and  $p_{\rm AR}/p_{\rm C}$ is taken to be at least four, although larger values are used in some special cases. Note that the  plasma density is calculated from the ideal gas law using the know pressure and temperature profiles once that $\alpha(x,y,z)$ and $\beta(x,y,z)$ have been obtained numerically.

\begin{figure}[h]
\center
\includegraphics[width=9cm]{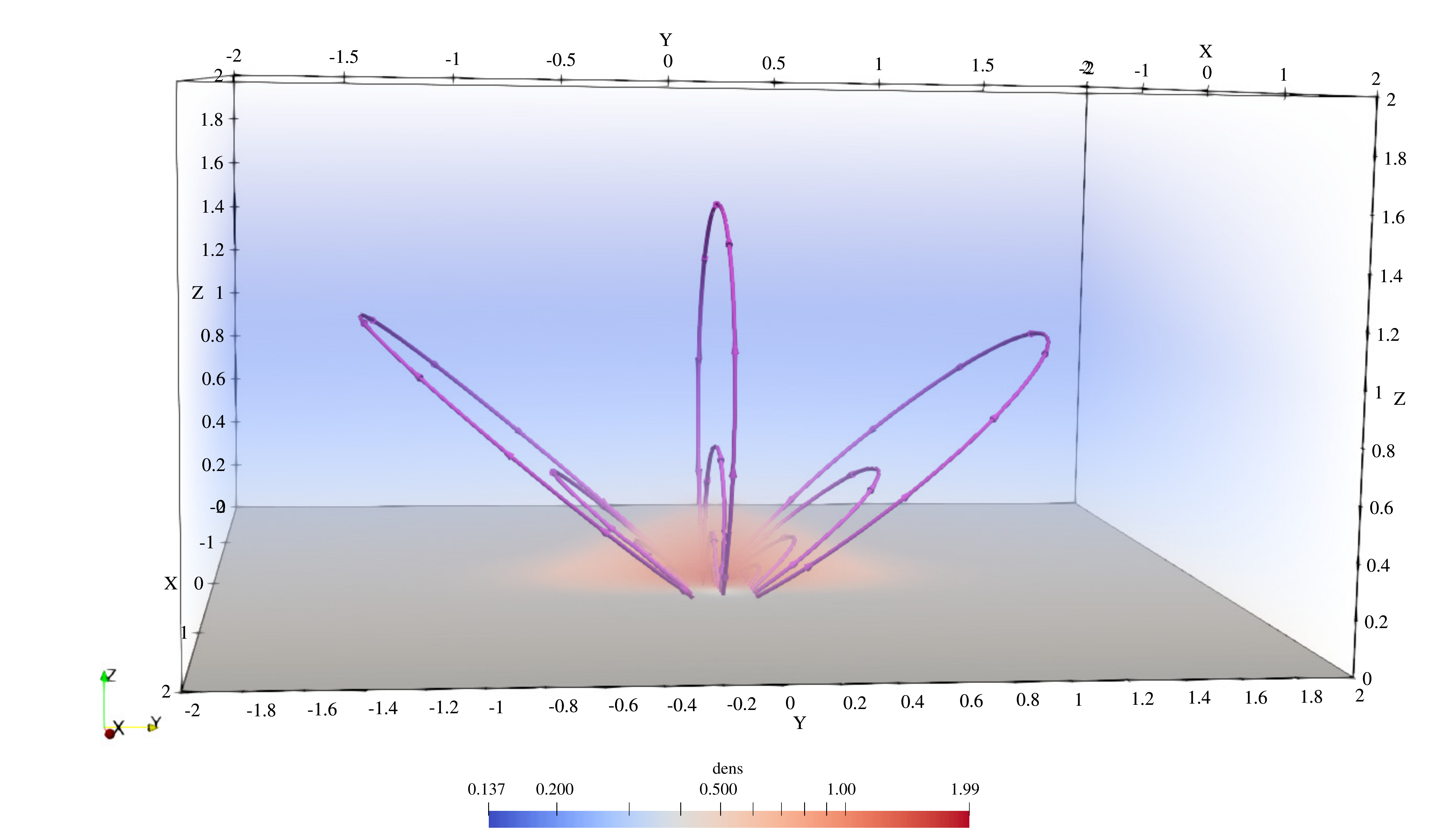}
\includegraphics[width=9cm]{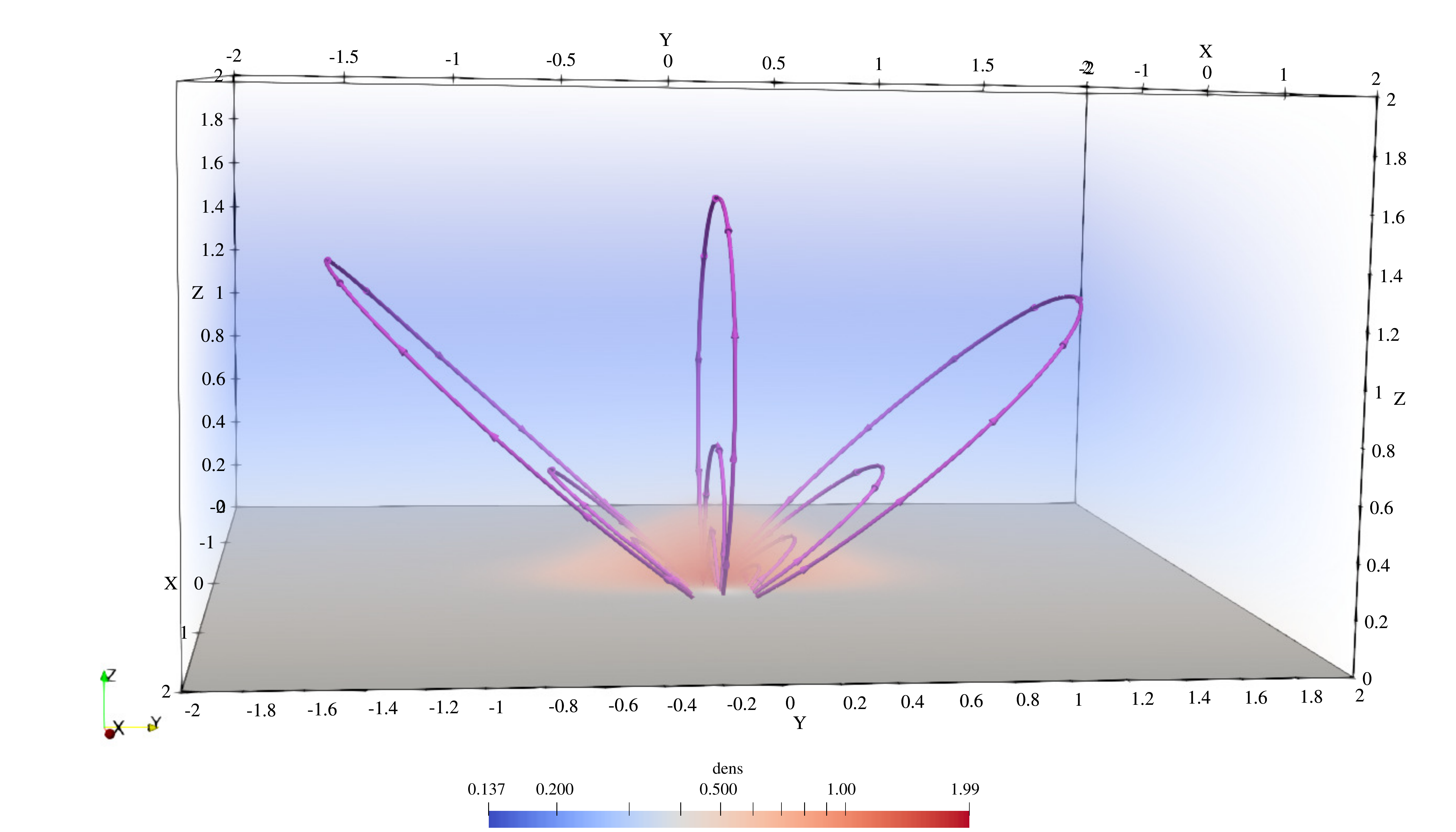}
\caption{\small Magnetic field lines and 
volume density
(normalised to the coronal density) for $\beta_{00}=1.4\times 10^{-3}$ (top panel) and  for  $\beta_{00}=1.4\times 10^{-1}$ (bottom panel). In this example $p_{\rm AR}=4 p_{\rm C}$, $T_{\rm AR}=2 T_{\rm C}$, $d=0.25$, $z_0=0.1$ and $\alpha_{\rm ref}=\alpha(0,0,0)$.} \label{fig:fieldexp}
\end{figure}

\subsection{Elementary example}

An example of an equilibrium numerically obtained with the previous pressure and temperature dependencies 
 on
$\alpha$ is shown in Fig.~\ref{fig:fieldexp} (top panel). We observe the bipolar distribution of the magnetic field but also the density concentration at the core of the AR. As the modified plasma-$\beta$ increases, the bipolar magnetic field expands to compensate the effect of a high pressure core. This effect is evident from the comparison of top panel and bottom panel of Fig.~\ref{fig:fieldexp}. The magnetic field lines that shown the largest displacements are located at the lobes of the bipolar region. This can be explained by
the interplay of the different forces. 
Consider the line that starts at $x=0$, $y=0$, $z=0$ and ends at $x=0$, $y=0$ and $z=z_{max}$. This line matches with the apex of all the magnetic field lines that are on the plane $y=0$. In the potential case at any point on the line there is a balance between the tension force pointing downward and the magnetic pressure force pointing upward. In the non-potential situation a new equilibrium is achieved on our reference line where the magnetic tension and the gravity force point downward while the total pressure force (gas plus magnetic) is pointing upward, 
balancing out
the total downward force. In comparison to the potential case the magnetic forces are weaker 
and  the magnetic field lines are displaced 
in the vertical direction. 
This last effect is more clear at the sides of 
the bipolar region where the gravity force is not 
aligned with the tension force as in our central reference line. 
In fact, these magnetic field lines are not coplanar 
although this feature is not 
visible in the plots.


From Fig.~\ref{fig:fieldexp} we also realise that the density enhancement across the core of the AR in the $x-$direction has a shorter spatial scale than that along the $y-$direction where the density is much more elongated. This is because $\alpha$ changes more rapidly in the $x$ than in the $y$ coordinate and this affects the pressure, the temperature and thus the density distribution of the AR. The rotational symmetry of the magnetic field also contributes to have elongated structures in the $y-$direction. 

\begin{figure}[h]
\center
\includegraphics[width=9cm]{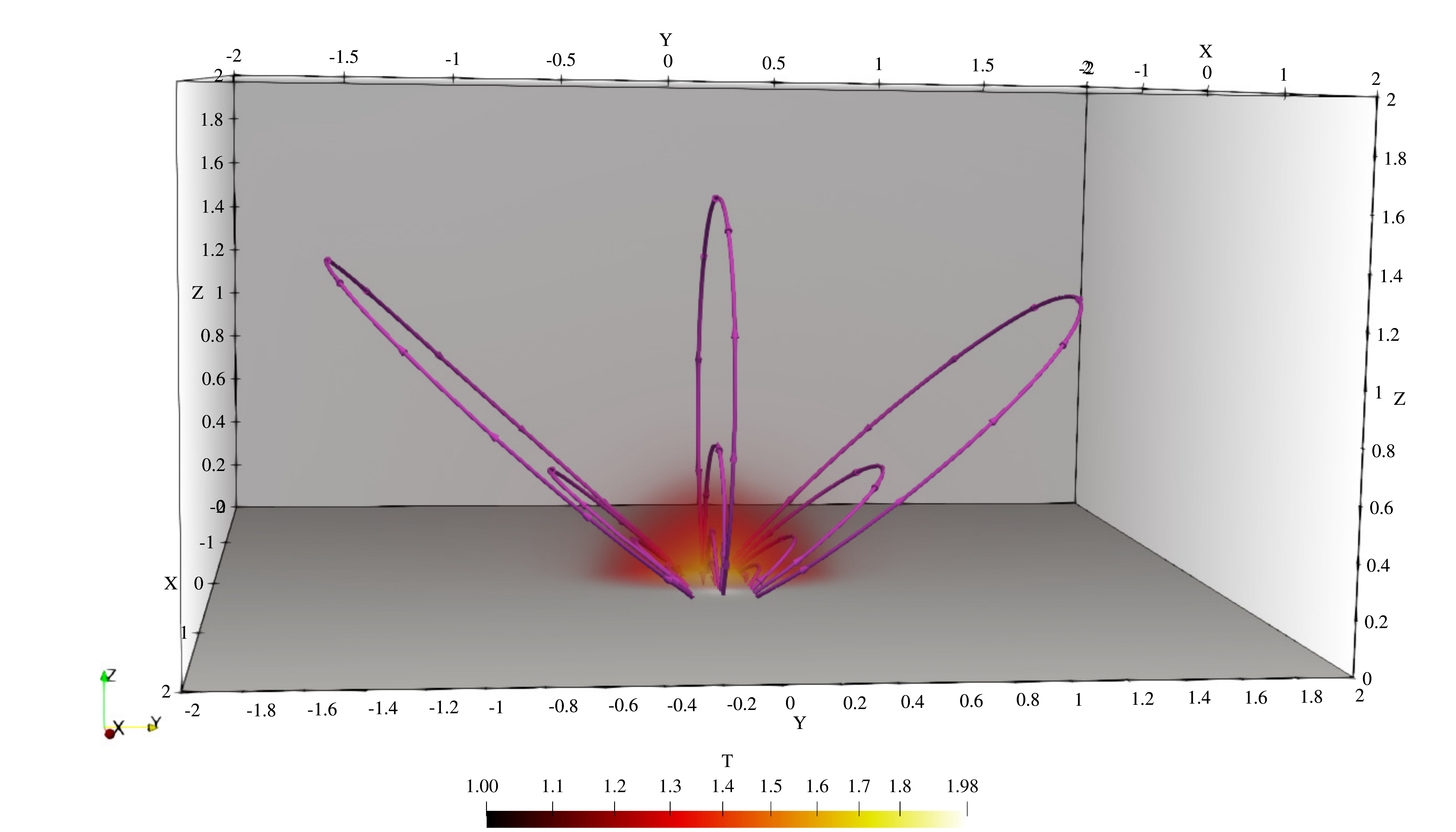}
\caption{\small Magnetic field lines and temperature volume for the same model as in Fig.~\ref{fig:fieldexp} bottom panel. Temperature is normalised to 1 MK.} \label{fig:fieldexptemp}
\end{figure}

The temperature distribution of the AR is displayed in
Fig.~\ref{fig:fieldexptemp}. 
It shows a hot core at 2 MK that smoothly 
matches with the coronal temperature of 1 MK of the environment. 
The spatial distribution of the temperature and the density
within the core of the AR are not exactly identical (compare
Fig.~\ref{fig:fieldexptemp} 
with Fig.~\ref{fig:fieldexp}). In particular, the temperature is
more localised towards the center of the AR.
We remind the reader 
that density is constructed 
from the ideal gas law using the assumed 
profiles for pressure and temperature which depend only on $\alpha$.

\begin{figure}[h]
\center
\includegraphics[width=10cm]{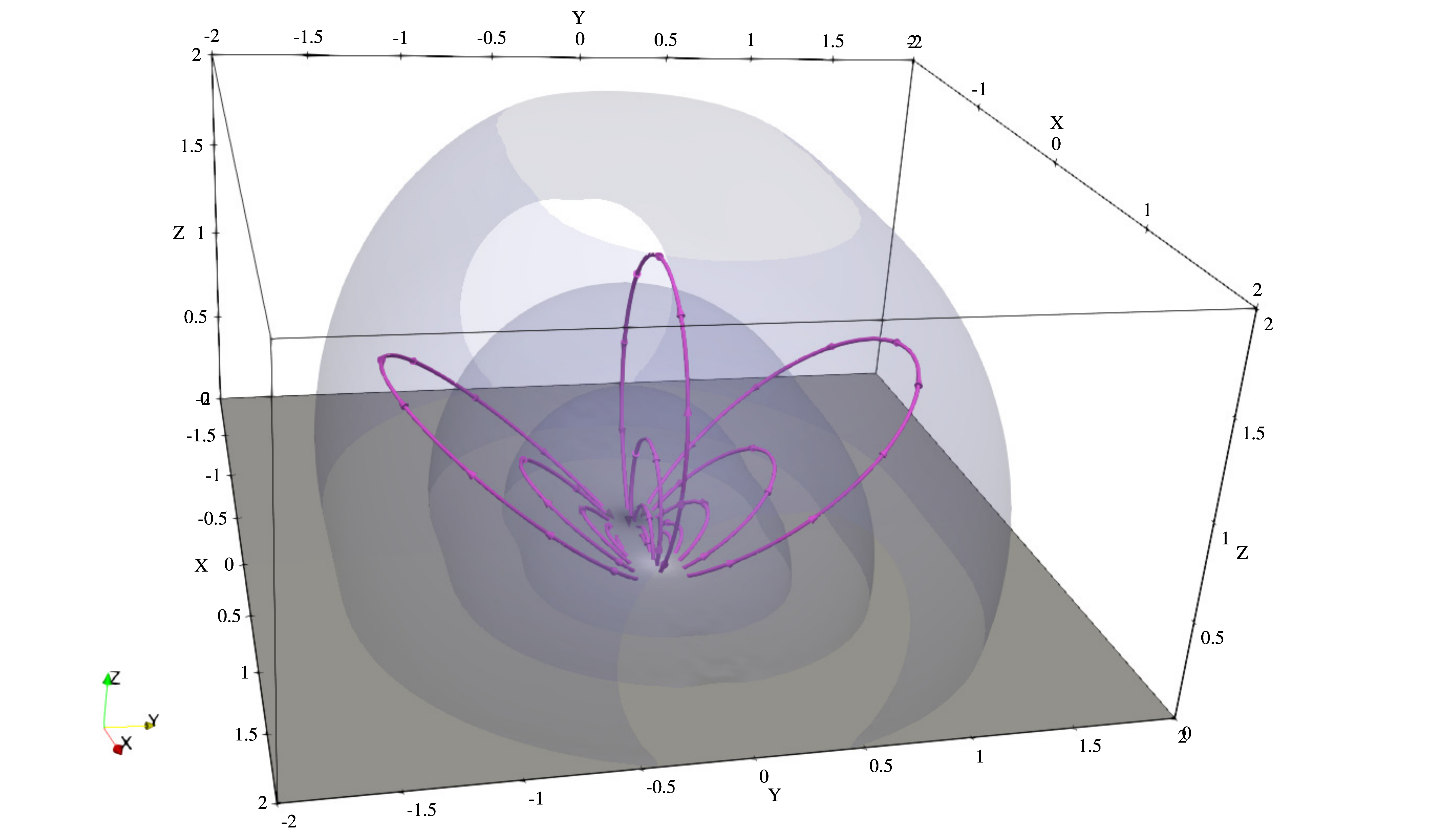}
\caption{\small Magnetic field lines and plasma-$\beta$ 
%
isosurfaces.
The three isosurfaces
represent values of 0.1, 0.01, 0.001, when moving from the exterior to the interior 
surfaces.
This example corresponds to the case shown in Fig.~\ref{fig:fieldexp} top panel.} \label{fig:fieldexpbet}
\end{figure}

The local plasma-$\beta$ is another quantity
of relevance for
the equilibrium. It is shown in Fig.~\ref{fig:fieldexpbet} as a set of concentric 
isosurfaces.
At the AR core this parameter is rather low, typically below 0.001 but as we move outwards the values continuously increase. Depending on the reference plasma-$\beta$ in the system, i.e., $\beta_{00}$, it may happen that it is well above one. This agrees with the inferred behaviour from observations. \citet{gary2001} found that although the plasma-$\beta$  is typically below one
in the corona, it takes values greater than one below the chromosphere (not included in our model) and moving from the photosphere upwards it can return to values around 1 at relatively low coronal heights, typically of the order of 1.2 solar radii. Therefore, values of the plasma-$\beta$ larger than one near the upper boundary of our domain 
should not
be considered as 
unrealistic.

To understand the effect of changing the plasma-$\beta$ in the system we have varied this parameter and calculated a set of equilibria using the symmetric bipole as boundary condition. For each equilibrium we have computed the different terms in the virial theorem, Eq.~(\ref{eq:virialn}), by numerically evaluating the surface integrals in 2D and the volumetric integrals in 3D. As mentioned earlier in Sect.~\ref{sec:method} the agreement of the numerical results with  the virial theorem is 
excellent, which
is an indication 
of the quality of
the 
numerical
solutions.

\subsection{Compound examples}\label{sec:shear}

Once we know the main characteristics of the simple symmetric bipolar magnetic field we explore other setups. We use the results presented in Sect.~\ref{analyticcalc} which allows us to construct a variety of models that might be of interest and that are used as boundary conditions. We begin by including magnetic shear and  as in the previous sections, we start with the reference initial potential solution and then the parameters $s_x$ and $s_y$ related to the affine shear transformation are gradually increased.

\begin{figure}[h]
\center
\includegraphics[width=12cm]{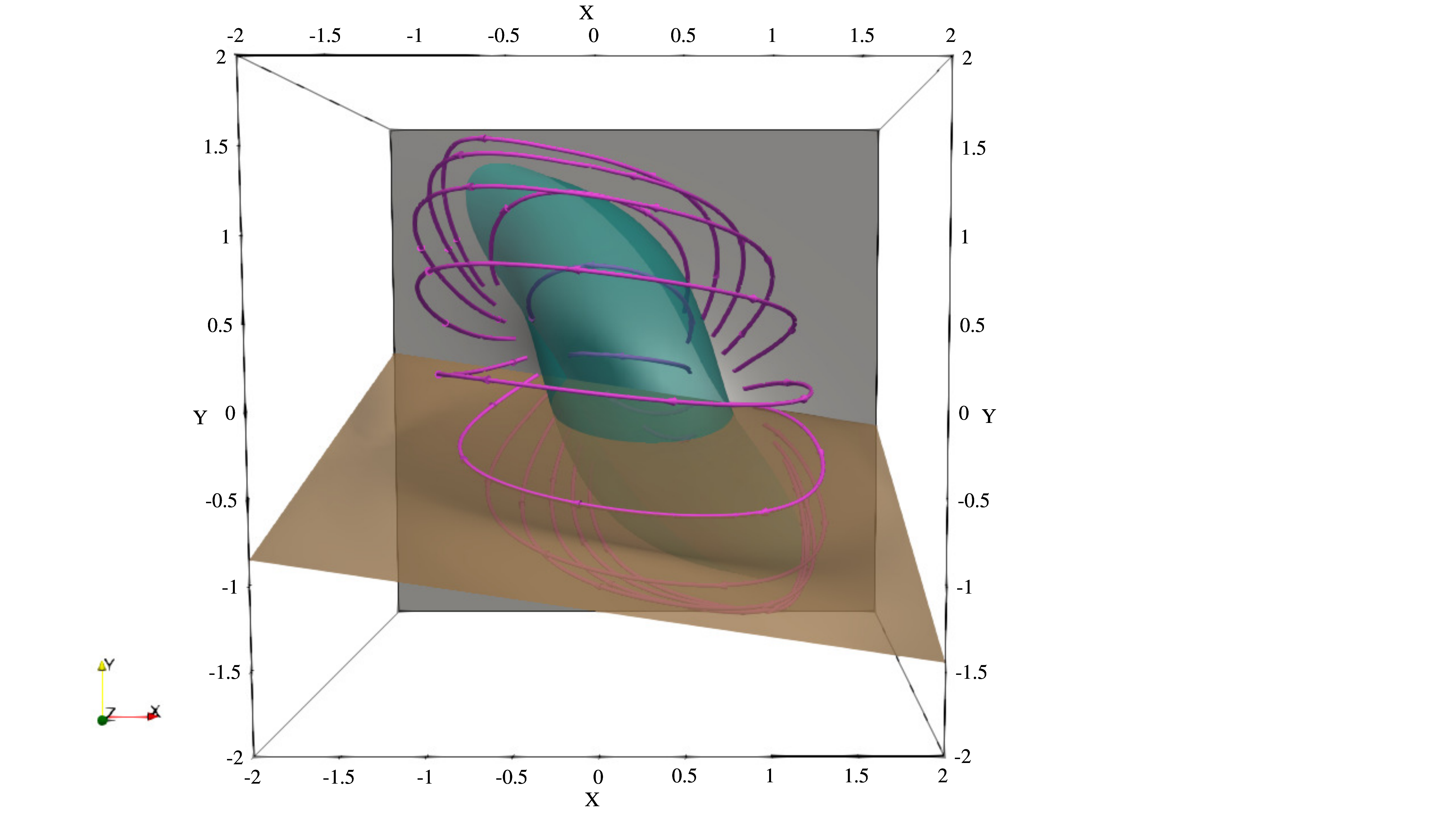}
\caption{\small Non-potential magnetic field lines and magnetic  isosurfaces calculated numerically by applying an affine shear transformation on the EPs with $s_y=0.6$ and $s_x=s_y/4$. Gas pressure and gravity are included. In this model the minimum plasma beta is around $1.4\times 10^{-3}$ while the maximum is around 2.5, well above one. In this example we use a bipole with $d=0.25$ and $z_0=0.5$ in the domain $x_{min}=-2$, $x_{max}=2$, $y_{min}=-2$, $y_{max}=2$, $z_{min}=0$ and $z_{max}=4$. } \label{fig:shearbet}
\end{figure}

An example of a three-dimensional sheared solution is shown in Fig.~\ref{fig:shearbet}. The deformations of the isosurfaces of the two EPs are evident and produced by the magnetic shear that can be seen at $z=0$ in the vertical distribution of the magnetic field (gray scale). The symmetry is lost in comparison to the potential case.
A  method 
to quantify the amount of shear in the configuration is to compute the integrated current, defined as $H_{\rm C} =\int_V {\bf B} \cdot {\bf J}\, dV.$
For a purely potential magnetic field we have that $H_{\rm C}=0$ (since ${\bf J}=0$). The current helicity is different from zero when the magnetic field is force-free (${\bf J} \times {\bf B}=0$ but ${\bf J}\neq 0$). We basically find a linear increase of $H_{\rm C}$ with $s_y$. The inclusion of gas pressure and gravity modifies the magnetic structure of the AR as well. There is a dense and hot plasma core in the present example (similar to that in Fig.~\ref{fig:fieldexp} but not symmetric, this is not 
shown in  Fig.~\ref{fig:shearbet}). Note how the intersection of the EP $\beta$ with the computational box is always a straight line, this is due to the applied BCs. But now the constant $\beta$ planes are inclined with respect to the sides of the box in the $y-$direction (compare for example with Fig.~\ref{fig:eulerpots}) as a consequence of the applied shear.

\begin{figure}[h]
\center
\includegraphics[width=12cm]{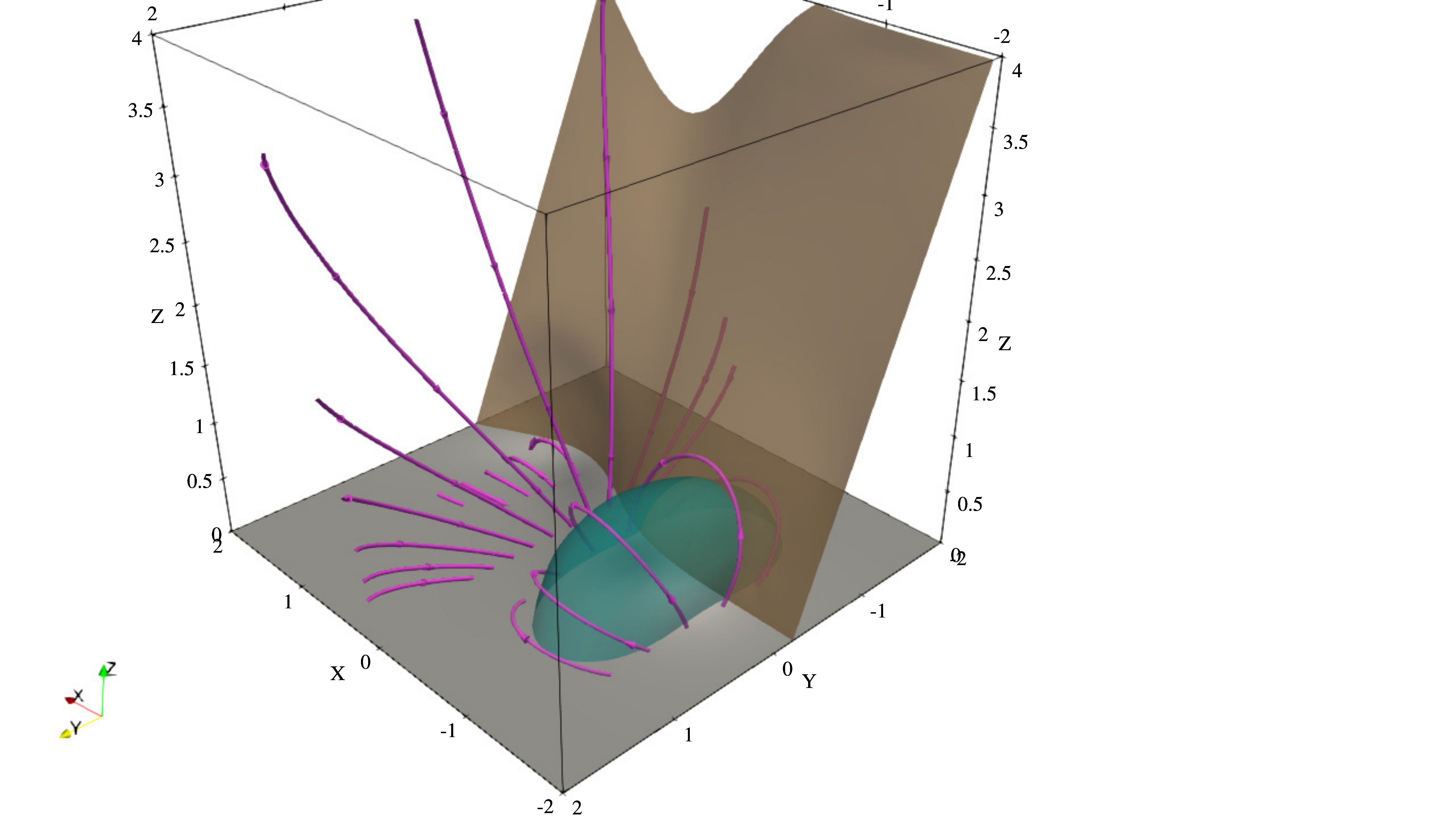}
\caption{\small Non-potential magnetic field lines and isosurfaces calculated numerically by applying a transformation on the EPs using Eq.~(\ref{eq:sheardispl}). Gas pressure and gravity are present. Compare with Fig.~\ref{fig:eulerpotnosymm}. In this configuration $a_0=0.6$, $x_0=1$, $w=1$, $\beta_{00}=1.4\times 10^{-3}$ and $\alpha_{\rm ref}=\alpha(-1,0,0)$.} \label{fig:sheartrans}
\end{figure}

We now 
apply
the transformation to
the EPs using the displacement of the polarities on the plane $z=0$ given by Eq.~(\ref{eq:sheardispl}). The obtained non-symmetric structure is shown in Fig.~\ref{fig:sheartrans}. This example indicates the presence of regions of open magnetic field close to one of the footpoints of the AR and it is the extension of the case shown in Fig.~\ref{fig:eulerpotnosymm}. This equilibrium could be used to model the interaction of an AR and a CH as was done in 2D in \citet{terradasetal2022}. The intersection of the EP $\beta$ with the upper and lower 
boundaries
of the computational box is not a straight line now and it is produced by the specific transformation used in this example. Interestingly, although not displayed in Fig.~\ref{fig:sheartrans}, the density is high 
within
the core of the ARs but low (below the coronal reference value) where the magnetic field lines are open. This behaviour is due to the dependence of gas pressure and temperature with $\alpha$ and agrees with the expected behaviour in CHs.

\begin{figure}
\center
\includegraphics[width=12cm]{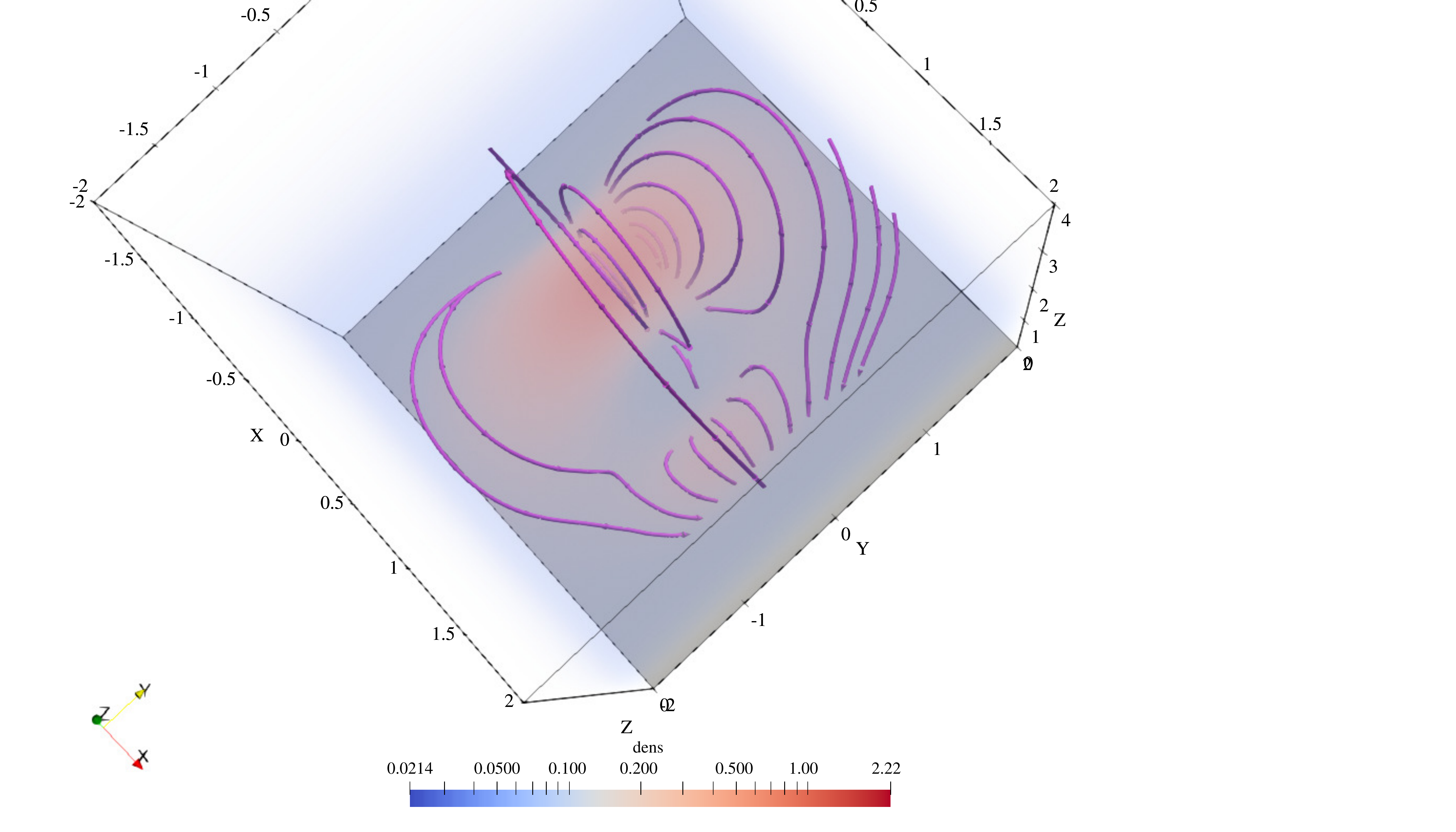}
\caption{\small Magnetic field lines and 
volume density (normalised to the coronal density)  for a situation that contains a null line. In this example magnetic shear with $a_0=0.3$, $x_0=1$, and $w=1$ has been induced in the structure. Gas pressure and gravity are present with $\beta_{00}=1.4\times 10^{-3}$.} \label{fig:eulerpotsbetgshearX}
\end{figure}

Finally, the density profile for a situation that contains a magnetic null line is shown in Fig.~\ref{fig:eulerpotsbetgshearX} and it is based on the example shown in Fig.~\ref{fig:eulerpotxpoint}. We have included a small amount of magnetic shear using the transformation applied in the previous example. We find a rather non-uniform distribution of the density, but most of the mass is located in the bipolar region around $x=-1$ where the plasma has the maximum temperatures because according to our model $\alpha^2$ has a maximum there. Around the null line the density and temperature are rather low. The numerical method is able to  converge to an equilibrium solution that contains both an ordinary AR and a nearby null line under the presence of gas pressure and gravity.

According to the different examples shown in this section we can affirm  that the EPs are useful to calculate a variety of magnetic structures, not necessarily simple, in which the plasma is coupled to the magnetic field. These configurations can be used to represent ARs in the solar atmosphere either as isolated entities or in companion of regions of open magnetic field or even to describe more involved magnetic topologies that contain null lines. 

\section{Stability analysis of the calculated equilibria}\label{sectinstability}

So far we have concentrated on the calculation and analysis of equilibria in 3D but their stability has not been addressed yet. A question that arises is whether the interplay between the magnetic field, gas pressure and gravity may lead to the appearance of 
magnetic Rayleigh-Taylor instabilities or Parker's instabilities, this is related to the second term in Eq.~(\ref{eq:energyprinciple}). There are two modes 
of the
magnetic buoyancy
instability, the undular mode with a wavenumber parallel to the magnetic field, and the interchange
mode with a wavenumber perpendicular to $\bf B$. The undular mode is typically named 
the Parker instability after \citet{parker1966}. The 
interchange mode is often referred to as the flute instability or the magnetic Rayleigh–Taylor instability. The undular mode occurs for long wavelength perturbations along the magnetic field lines
and
the gas tends to slide downward along the field from the peaks into the valleys further enhancing the undulations.
On the 
other hand,
the interchange mode occurs for short 
wavelength perturbations, when the interchange of two straight
flux tubes reduces the potential energy in the system. Nevertheless, the most unstable mode has a 3D structure, and therefore has both a wavevector component parallel and perpendicular to the magnetic field. The instability typically occurs for short wavelengths in the transverse direction but it is highly dependent on the wavelength parallel along the field, being completely suppressed for short parallel wavelengths \citep[see for example, Fig.~13.1 in][]{parker1979}.

In general, an equilibrium is stable if and only if the change in the potential energy, $\delta W$, associated to all allowable displacements satisfying appropriate boundary conditions is always positive, i.e. $\delta W\ge 0$. When the equilibrium configuration is complex (three-dimensional curved magnetic field in balance with the pressure gradient and the gravity force) the estimation of the sign of the potential energy must be inevitably done by numerical means. We give further details about a numerical approach that has received little attention in the literature but that it is of considerable utility.
\citet{zwingmann1984,zwingmann1987} \citep[see also][]{schindler2006} generalised the stability criteria found by \citet{schindleretal1983} and \citet{hood1984}
to three dimensions and showed that the potential energy can be split  into two parts
\begin{align}\label{eq:energyprinciple}
    \delta W=\delta Q+\delta^2F.
\end{align}
The term $\delta Q$ accounts for the purely convective instability and a stable situation  is achieved when \citep[see for example][]{schindleretal1983}
\begin{align}
    -\frac{R}{\mu g} \frac{\partial T}{\partial z}\leq \frac{\gamma-1}{\gamma}. \label{eqSchwartz}
\end{align}
This condition is the 
Schwarzschild
criterion for stability against convection modes projected along the magnetic field lines. Since in our case according to Eq.~(\ref{eqpressalongBs}) there is no explicit dependence of temperature 
upon
$z$ (but note that $\alpha$ and $\beta$ depend on $x$, $y$ and also $z$) the previous condition is always satisfied and the system is convective stable.

The second term in Eq.~(\ref{eq:energyprinciple}) can be written as
\begin{align}
    \delta^2 F= \frac{1}{2\mu_0}\int_V (\delta \alpha^*,\delta \beta^*) D \left(\begin{matrix}
    \delta \alpha \\
    \delta \beta
\end{matrix}\right) dV,\label{eq:functional}
\end{align}
where $D$ is a linear operator written as a $2\times 2$ matrix, where we have defined $\delta \alpha=-{\boldsymbol{\xi}} \cdot \nabla \alpha$, $\delta \beta=-{\boldsymbol{\xi}} \cdot \nabla \beta$, and 
${\boldsymbol{\xi}}$ represents the displacement vector. We construct the matrix $M$,
which is the numerical equivalent of
the linear MHD operator $D$ when we perform a discretization using finite elements.  From the mathematical point of view the matrix $M$ is positive definite if ${\bf x}^T M {\bf x}>0$ for any vector ${\bf x}$ (being ${\bf x}^T$ its transpose) and this is in essence the integrand in Eq.~(\ref{eq:functional}). Therefore,   if $M$ is positive definite then the equilibrium is stable and any changes of $M$ from positive definite to non-definite
indicate a transition 
from
stable to unstable solutions because change the sign of $\delta^2 F$. This is due to the fact that this matrix is 
the numerically discretised equivalent of
a sufficient stability functional (when purely convective instabilities are not present in the system) and it turns out that $\delta^2F$ is the second variation of the free-energy functional introduced by \citet{grad1964}. 
This method based on the analysis of the matrix $M$ has been successfully used in 2D studies \citep[e.g.][]{zwingmann1987,plattneukirch1994,neukirchromeou2010}. If the configuration is three-dimensional this approach is still valid but involves significantly larger matrices than in the 2D case and therefore more computationally intensive calculations. The disadvantage of the method is that it does not provide the frequency or the growth rates of the unstable modes. It just classifies the system as stable or unstable, which is nevertheless a valuable piece of information.

The procedure is clear now, after obtaining a solution to the equilibrium equations, the matrix $M$ associated to the linear operator is constructed and its positiveness is evaluated to assess the stability of the solution. \textsc{Mathematica} has implemented a command to test if a matrix is positive definite. In the work of \citet{zwingmann1987} the linearised operator is also needed to solve the equilibrium equations but in our case it is better to construct the matrix $M$ a posteriori once we have computed the equilibrium. Details about the linear operator are given in Appendix B.


\subsection{Parker's instability test: horizontal magnetic field}

As a test of the numerical procedure based on the matrix $M$ we have considered the simplified case of a purely horizontal magnetic field coupled to gas pressure.  Pressure, density and magnetic field decay exponentially with height but the sound and Alfv\'en speeds are constant, allowing a normal model analysis. The details about the stability analysis of this elementary configuration can be found in \citet{parker1966,parker1979}. The general dispersion relation is given by Eq.~(13.33) of \citet{parker1979} and depending on the choice of the wavenumbers in the three directions, $k_x$, $k_y$ and $k_z$, and the value of the plasma-$\beta$ (constant in the domain) the frequency switches from purely real to purely imaginary meaning that the system changes from stable to an unstable state. 

We have derived the EPs associated to the previous equilibrium, with the magnetic field pointing the $x-$direction, which read
\begin{align}
    \alpha&=2 B_0\, \mathcal{H}\, e^{-z/(2\mathcal{H})},\\
    \beta&=y,
\end{align}
while gas pressure is
\begin{align}
    p&=p_{00} e^{-z/\mathcal{H}}=p_{00} \left(\frac{\alpha}{2 B_0 \mathcal{H}}\right)^{2\left(1-\mathcal{H}/H\right)} e^{-z/H},
\end{align}
and has been written in terms of the EP $\alpha$. Here, $\mathcal{H}=H+H/\beta_{00}$ is the modified pressure scale height that couples the plasma to the magnetic field ($H$ is the isothermal scale height). The key parameters of this problem are $\beta_{00}$ and the wavelengths that fit in our computational box, namely,
\begin{align}
    k_x&=2 \pi/(2(x_{max}-x_{min})),\label{eq:kx}\\
    k_y&=2 \pi/(2(y_{max}-y_{min})),\label{eq:ky}\\
    k_z&=2 \pi/(2(z_{max}-z_{min})).\label{eq:kz}
\end{align}
Using the previous expressions we have tested the stability of the system using the positive definiteness
of the matrix $M$ constructed using finite elements. We find that the
positive definiteness
of $M$ changes precisely when the dispersion relation given by Parker predicts a transformation of a stable solution into an unstable solution, or vice versa. This match indicates that the numerical method is correctly implemented and can be confidently used to investigate 
the stability of
more complex equilibria. Small deviations from the theoretical predictions are related to the fact that in our calculations we are imposing strict line-tying conditions while in Parker's calculations Fourier analysis 
without imposing line-tying
was performed.

\subsection{Results for the computed 3D equilibria}

\begin{figure}
\center
\includegraphics[width=8cm]{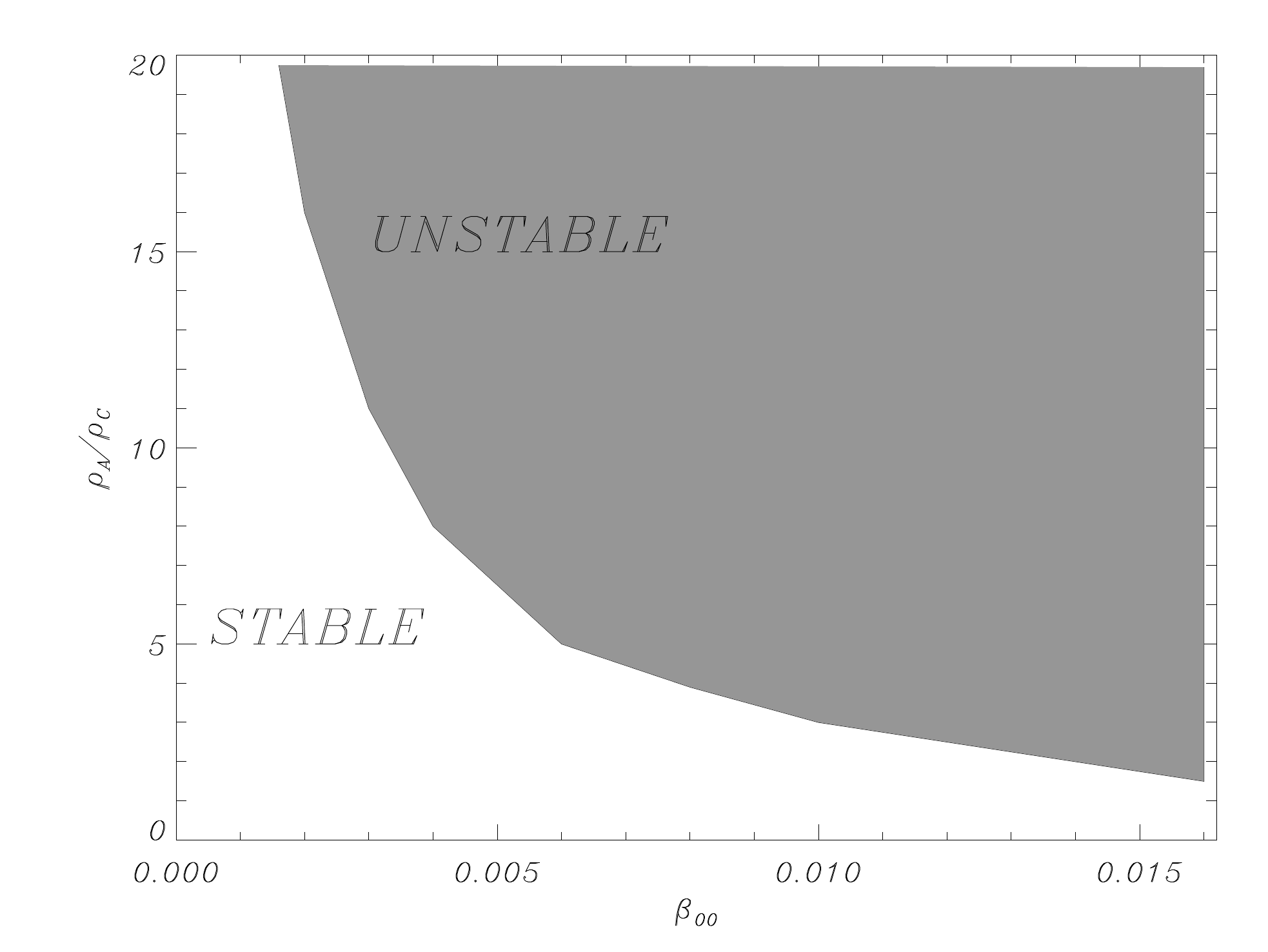}
\caption{\small Stability diagram for a bipolar configuration. The gray region corresponds to unstable solutions while the white region to stable states. The vertical axis represents the density contrast between the core of the AR and the external environment, while the horizontal axis corresponds to the reference plasma-$\beta$, $\beta_{00}$. In this plot we have used a bipole with $d=0.25$ and $z_0=0.5$ in the domain $x_{min}=-2$, $x_{max}=2$, $y_{min}=-2$, $y_{max}=2$, $z_{min}=0$ and $z_{max}=4$. The temperature contrast, $T_{\rm AR}=2 T_{\rm C}$, is the same for all the points. } \label{fig:stability}
\end{figure}

When the reference plasma-$\beta$ is very small
 most of the numerically calculated equilibria described in Sect.~\ref{sect:numequil} are found to be stable. However, we have found that increasing this parameter and changing the pressure contrast (density contrast between the AR and the environment) may lead to unstable structures. An example of a stability diagram based on the bipole considered in Sect.~\ref{subsect:bipole} is shown in Fig.~\ref{fig:stability}. For a fixed reference plasma-$\beta$, $\beta_{00}$, we always find that by increasing the density contrast at the core of the AR the configuration becomes unstable at some point, meaning that a small perturbation in the system will automatically lead to the growth of the perturbation. The magnetic field cannot support,  under stable conditions, the extra 
 mass 
 load
 due to the increase of the density contrast. When the plasma-$\beta$ is very small and therefore the magnetic field is very strong in comparison to the gas pressure, increasing the amount of mass at the core of the AR  makes the model unstable only for very large and unrealistic density contrasts according to Fig.~\ref{fig:stability}. This behaviour is
 as it would be expected
 from the physical point of view.
 
 The stable/unstable 
 parameter regions
 also depend on the size of the computational box since this parameter changes the wavenumbers that fit into the domain (see Eqs.~(\ref{eq:kx})-(\ref{eq:kz})). Starting from a stable situation and increasing the length of the domain in each direction inevitably leads to an unstable situation because the maximum wavelengths increase. In particular the wavelength along the magnetic field increases because the field lines are longer in the 
 larger box.  The opposite is also true, we can start with an unstable state but by progressively reducing the size of the domain a stable situation is eventually achieved. This is in qualitative agreement with the behaviour found in Parker's problem and it is a characteristic feature of the magnetic buoyancy instability. 
 
 We have also investigated the effect of shear on the stability of the configurations. We have not reported the suppression of the instability by magnetic shear as one would expect for purely interchange modes. This is an additional indication that the instability found in this work is dominated by to the undular component of the 3D modes. Nevertheless, we have not considered magnetic structures with a large amount of shear that could have a strong effect on the characteristics of the modes.

\section{Discussion and conclusions}

We have carried out a first exploratory study of the application of EPs in 3D to obtain MHS solutions under the presence of gas pressure and gravity that can represent ARs. Due to the nonlinearity of the equilibrium equations one of the main difficulties of the use of EPs \citep[see][]{stern1970}  
is
the 
calculation
of these variables even when the magnetic field is known. Nevertheless, we have shown that using the potential solution for the magnetic field as a starting point, the EPs can be numerically calculated when departures from the current-free case are considered. In particular, we have used as initial distributions of $\alpha$ and $\beta$ analytical expressions based on an axially symmetric  potential configuration. Besides the inclusion of the pressure gradient force and the gravitational force we have additionally incorporated shear in the magnetic field. We have investigated different methods and have found that the most efficient procedure is to gradually transform the potential magnetic field at the boundaries of the domain into a non-potential field by applying  specific transformations that naturally increase the current helicity in the system. This method is effective and transparent and can be viewed as a particular case of other families of transformations that could be investigated in the future. 

The choice of the particular functional dependence of gas pressure and temperature 
on
the EPs determines the type of thermal structure of the models. This choice is in principle arbitrary but based on the typical features of ARs, with hot and dense plasma cores, we have shown that assuming that pressure and temperature are proportional to the square of $\alpha$ \citep[see also][]{terradasetal2022} we find that we reproduce these basic 
features
of ARs. It is worth noting that $\alpha$ and $\beta$ can be gauged and therefore the thermal structure changes according to the specific gauge. Constraints on these gauges need to be imposed based on the observational data. Our models describe the main features of ARs regarding the diffuse background and in any case they do not intend to explain the fine structure of coronal loops embedded in ARs which are most likely due to localised heating.

The three-dimensional configurations under force balance that we have numerically computed can be used in the future to study the propagation and interaction of global MHD waves with ARs, i.e. can be used in time-dependent simulations. The interplay of global MHD waves with CHs, poorly addressed in the literature due to the lack of 3D models, can be also investigated using the approach proposed in the present paper. 

Although the topology of the magnetic field must be simple to use a description based on EPs (see Sect.~\ref{sectintro}) we have shown that it is possible to obtain MHS solutions that include null lines under the presence of gas pressure and gravity. These critical points can still be represented using EPs although we have not investigated geometries with truly 3D null points that may contain  
spine lines and fan planes, and where the description based on EPs fails \citep[see][]{hesseschindler1988}. Even with this limitation, studies about the propagation of waves around null lines can benefit from the models proposed here. 

We have provided an
initial investigation
of the stability properties of the numerically computed equilibrium configurations
by the application of the sufficient stability criterion established by \citet{zwingmann1984,zwingmann1987}, 
but here applied in 3D. An 
interesting property
of the constructed models is that for some values of the parameters, namely, high plasma-$\beta$, dense active region cores and significantly big spatial domains, the system is unstable.  For these unstable modes the magnetic buoyancy force dominates over the stabilising effects due to magnetic curvature, magnetic shear and line-tying, at least in the regime considered in the present work.

Interestingly, the formalism used in the current paper can be extended to the case with pressure and temperature (and therefore density) changing explicitly with height. This could be used to include, for example, the effect of an idealised chromosphere in the  model through the presence of a low temperature and a high density layer. Hence, the method used in this work could be applied not only to the solar corona but also to lower layers in the solar atmosphere, as it has been recently done using other approaches by \citet{zhuwiegelmann2018,zhuwiegelmann2022}. In relation to this  problem,  it is important to mention that the instability criterion applied in this work is only valid when no purely convective instabilities are present. The isothermal case along the magnetic field lines studied in this paper fits into this category, but this does not need to be the case in a  more realistic
situation, specially if the transition between the chromosphere and corona is included in the models. Nevertheless, the study of the stability of an arbitrary non-isothermal 3D configuration is 
a difficult
task that inevitably requires the calculation of the full spectrum of modes through the computation of the MHD eigenmodes. These computations are even more complex when the effect of flows are included in the models. This is a more realistic situation than the static case since there 
is ubiquitous evidence 
in the observations of the presence of flows in the solar corona and specially in ARs. As a first step we could suppose that the flow is field aligned and apply a similar formalism as the one used in the present paper by including additional terms in the equations that account for the flow effect \citep[see further details in Appendix 1 of][]{schindler2006}. 

Finally, it is important to emphasise that the families of equilibria  numerically constructed in the present paper cannot be
expected to model realistic solar ARs in
detail. Our results are useful as
direct demonstration of basic physical effects and as idealised situations, but they can still provide some physical insight into more realistic situations. For example, the energy considerations in the AR (radiative losses, thermal conduction and heating) and the fine structure due to magnetic loops have been completely ignored. But even with the limitations of the approach used in the present paper we think that the utilisation of EPs to describe ARs is 
useful and 
should be investigated in more detail in the future.

\begin{acknowledgements}
This publication is part of the R+D+i project PID2020-112791GB-I00, financed by MCIN/AEI/10.13039/501100011033. T. N. acknowledges financial support by the UK's Science and Technology Facilities Council (STFC) via Consolidated Grants ST/S000402/1 and ST/W001195/1.
The authors thank the anonymous referee for useful comments and suggestions that helped to improve the paper.
\end{acknowledgements}



\begin{appendix}

\section{The divergence form of the operator}

The equilibrium equations given by the PDEs in Eqs.~(\ref{eqFacrossa})-(\ref{eqFacrossb}) are written in divergence form. It is not difficult to show that using the divergence and the gradient operators the equations are 
\begin{align}
    \nabla \cdot ( -C\, \nabla \alpha + D\,\nabla \beta)&= -\mu_0\, \partial_{\beta} p, \nonumber \\
    \nabla \cdot ( -E\, \nabla \alpha +C\,\nabla \beta)&= \mu_0\, \partial_{\alpha} p,     \label{eqFacrossdiv}
\end{align}
where $C$, $D$, and $E$ are $3\times 3$ diagonal matrices given below. The compact form of the PDEs has some advantages if instead of Dirichlet conditions Neumann boundary conditions are used and is in general required when the problem is written in weak form (typically necessary when using finite elements).

The matrices that appear in the divergence form of  Eq.~(\ref{eqFacrossdiv}) are
\begin{align}
C=
\begin{pmatrix}
\partial_y\alpha\,\partial_y\beta +\partial_z\alpha\,\partial_z\beta& 0 & 0\\
0 & \partial_x\alpha\,\partial_x\beta +\partial_z\alpha\,\partial_z\beta & 0\\
0 & 0 &\partial_x\alpha\,\partial_x\beta +\partial_y\alpha\,\partial_y\beta
\end{pmatrix},
\end{align}
\begin{align}
D=
\begin{pmatrix}
(\partial_y\alpha)^2 +(\partial_z\alpha)^2& 0 & 0\\
0 & (\partial_x\alpha)^2 +(\partial_z\alpha)^2& 0\\
0 & 0 &(\partial_x\alpha)^2+(\partial_y\alpha)^2
\end{pmatrix},
\end{align}
\begin{align}
E=
\begin{pmatrix}
(\partial_y\beta)^2 +(\partial_z\beta)^2& 0 & 0\\
0 & (\partial_x\beta)^2 +(\partial_z\beta)^2& 0\\
0 & 0 &(\partial_x\beta)^2+(\partial_y\beta)^2
\end{pmatrix}.
\end{align}

\section{The divergence form of the linearised operator}

The linear operator is derived by 
linearising Eq.~(\ref{eqFacrossdiv}), i.e.\ by assuming that $\alpha'$ and $\beta'$ are small perturbations on the initial equilibrium vales $\alpha$ and $\beta$. Using the Taylor expansion of gas pressure around the equilibrium and keeping 
terms 
up to first order in the perturbed quantities, 
the result, written again in divergence form for completeness, is
\begin{align}
    -\nabla \cdot (C\nabla \alpha')-\mu_0 \partial^2_{\alpha \alpha}p \alpha' + \nabla \cdot (\bar{D}\nabla \beta')+{\bf J} \cdot \nabla \beta'-\mu_0 \partial^2_{\alpha \beta}p \beta' &= 0, \nonumber \\
    \nabla \cdot ( D \nabla \alpha') -{\bf J} \cdot \nabla \alpha'-\mu_0 \partial^2_{\alpha \beta}p \alpha'- \nabla \cdot (\bar{C}\nabla \beta')-\mu_0 \partial^2_{\beta \beta}p \beta'&=0,     \label{eqFacrossdivlin}
\end{align}
where $\bf J$ is the equilibrium current density given by Eq.~(\ref{eqJEuler}), while
\begin{align}
C=
\begin{pmatrix}
(\partial_y\beta)^2 +(\partial_z\beta)^2& -\partial_y\beta\, \partial_x\beta & -\partial_z\beta\, \partial_x\beta\\
-\partial_y\beta\, \partial_x\beta & (\partial_x\beta)^2 +(\partial_z\beta)^2& -\partial_y\beta\, \partial_z\beta\\
 -\partial_z\beta\, \partial_x\beta & -\partial_y\beta\, \partial_z\beta &(\partial_x\beta)^2+(\partial_y\beta)^2
\end{pmatrix},
\end{align}
and 
\begin{align}
D=
\begin{pmatrix}
\partial_y\alpha\,\partial_y\beta +\partial_z\alpha\, \partial_z\beta& -\partial_y\alpha\, \partial_x\beta & -\partial_z\alpha\, \partial_x\beta\\
-\partial_x\alpha\, \partial_y\beta & \partial_x\alpha\, \partial_x\beta +\partial_z\alpha\, \partial_z\beta & -\partial_z\alpha\, \partial_y\beta\\
 -\partial_x\alpha\, \partial_z\beta & -\partial_y\alpha\, \partial_z\beta &\partial_x\alpha\,\partial_x\beta+\partial_y\alpha\,\partial_y\beta
\end{pmatrix}.
\end{align}

\noindent The matrix $\bar {C}$ is equal to $C$ but with the substitution of $\beta$ by $\alpha$, while $\bar {D}$ is the same
as $D$,
but with $\alpha$ and $\beta$ 
swapped. 
The non-divergence form of the linear operator is given in Eq.~(2.5) of \citet{zwingmann1987}.

\end{appendix}


\end{document}